\documentclass[longauth]{aa}  

\usepackage{amsmath}
\usepackage{graphicx}
\usepackage{txfonts}
\usepackage{natbib}

\bibpunct{(}{)}{;}{a}{}{,}

\def\teff{\ifmmode T_{\rm eff} \else $T_{\mathrm{eff}}$\fi}

\def\ltsima{$\buildrel<\over\sim$}
\def\lsim{\lower.5ex\hbox{\ltsima}}

\newcommand{\hii}{H~{\sc ii}}
\newcommand{\ha}{\ifmmode {\rm H}\alpha \else H$\alpha$\fi}
\newcommand{\hb}{\ifmmode {\rm H}\beta \else H$\beta$\fi}
\newcommand{\lya}{\ifmmode {\rm Ly}\alpha \else Ly$\alpha$\fi}

\newcommand{\ebv}{\ifmmode E_{\rm B-V} \else $E_{\rm B-V}$\fi}
\newcommand{\av}{\ifmmode A_{\rm V} \else $A_{\rm V}$\fi}
\def\micron{$\mu$m}
\def\kms{km s$^{-1}$}
\def\kmsmpc{km s$^{-1}$ Mpc$^{-1}$}

\def\msun{\ifmmode M_{\odot} \else M$_{\odot}$\fi}
\def\msunyr{\ifmmode M_{\odot} {\rm yr}^{-1} \else M$_{\odot}$ yr$^{-1}$\fi}
\def\zsun{\ifmmode Z_{\odot} \else Z$_{\odot}$\fi}

\def\lsun{\ifmmode L_{\odot} \else L$_{\odot}$\fi}

\def\mup{\ifmmode M_{\rm up} \else M$_{\rm up}$\fi}
\def\mlow{\ifmmode M_{\rm low} \else M$_{\rm low}$\fi}

\def\aap{A\&A}

\def\apjl{ApJL}
\def\apjs{ApJS}

\newcommand{\oh}{\ifmmode 12 + \log({\rm O/H}) \else$12 + \log({\rm
O/H})$\fi}
\def\flyf{\ifmmode f_{\rm Lyf} \else $f_{\rm Lyf}$\fi}
\def\pz{\ifmmode P(z) \else $P(z)$\fi}
\def\ki2{\ifmmode \chi^2 \else $\chi^2$\fi}
\def\zphot{\ifmmode z_{\rm phot} \else $z_{\rm phot}$\fi}

\newcommand{\xphot}{\ifmmode x_\gamma \else $v_\gamma$\fi}
\newcommand{\xobs}{\ifmmode x_{\rm obs} \else $x_{\rm obs}$\fi}
\newcommand{\xcmf}{\ifmmode x_{\rm CMF} \else $x_{\rm CMF}$\fi}
\newcommand{\vexp}{\ifmmode V_{\rm exp} \else $V_{\rm exp}$\fi}
\newcommand{\vmax}{\ifmmode V_{\rm max} \else $V_{\rm max}$\fi}
\newcommand{\nh}{\ifmmode N_{\rm HI} \else $N_{\rm HI}$\fi}
\newcommand{\dv}{\ifmmode \Delta v({\rm em-abs}) \else $\Delta v({\rm em}-{\rm abs})$\fi}

\def\fesc{\ifmmode f_{\rm esc} \else $f_{\rm esc}$\fi}

\def\frellya{\ifmmode f^{\rm rel}_{\rm{Ly}\alpha} \else $f^{\rm rel}_{\rm{Ly}\alpha}$\fi}

\def\ewlya{$EW({\rm{Ly}\alpha})$}

\def\hii{H{\sc ii}}

\newcommand{\mstar}{\ifmmode M_\star \else $M_\star$\fi}
\newcommand{\muv}{\ifmmode M_{1500} \else $M_{1500}$\fi}
\newcommand{\auv}{\ifmmode A_{\rm UV} \else $A_{\rm UV}$\fi}
\newcommand{\luv}{\ifmmode L_{\rm UV} \else $L_{\rm UV}$\fi}
\newcommand{\lir}{\ifmmode L_{\rm IR} \else $L_{\rm IR}$\fi}
\newcommand{\lbol}{\ifmmode L_{\rm bol} \else $L_{\rm bol}$\fi}
\newcommand{\liruv}{\ifmmode L_{\rm IR+UV} \else $L_{\rm IR+UV}$\fi}
\newcommand{\liroveruv}{\ifmmode L_{\rm IR}/L_{\rm UV} \else $L_{\rm IR}/L_{\rm UV}$\fi}
\newcommand{\nlyc}{\ifmmode N_{\rm Lyc} \else $N_{\rm Lyc} $\fi}
\newcommand{\rholyc}{\ifmmode \rho_{\rm Lyc} \else $\rho_{\rm Lyc} $\fi}
\newcommand{\chion}{\ifmmode \xi_{\rm ion} \else $\xi_{\rm ion}$\fi}
\newcommand{\chioncorr}{\ifmmode \xi_{\rm ion}^0 \else $\xi_{\rm ion}^0$\fi}

\newcommand{\cii}{\ifmmode {\rm [CII]} \else [C~{\sc ii}]\fi}
\newcommand{\lcii}{\ifmmode L({\rm [CII]}) \else $L({\rm [CII]})$\fi}
\newcommand{\Cii}{[C~{\sc ii}] 158 $\mu$m}
\newcommand{\citedelooze}{\cite{De-Looze2014The-applicabili}}
\newcommand{\citeharikane}{\cite{Harikane2019Large-Populatio}}
\newcommand{\citesurvey}{\cite{Le-Fevre2019The-ALPINE-ALMA}}
\newcommand{\citedata}{\cite{Bethermin2020The-ALPINE-ALMA}}
\newcommand{\citeancillary}{\cite{Faisst2020The-ALPINE-ALMA}}
\newcommand{\citecassata}{\cite{Cassata2020The-ALPINE-ALMA}}

\begin{document}

    \title{The ALPINE-ALMA \cii\ survey:}
  \subtitle{Little to no evolution in the \cii--SFR relation over the last 13 Gyr}

\author{D. Schaerer\inst{1,2}
\and M. Ginolfi\inst{1}
\and M. B\'ethermin \inst{3}
\and Y. Fudamoto \inst{1}
\and P.A.~Oesch \inst{1}
\and %
O.~Le F\`evre\inst{3}
\and A.~Faisst\inst{4}
\and P.~Capak\inst{4,5,6}
\and P.~Cassata\inst{7,8}
\and J.D.~Silverman\inst{9,10}
\and Lin Yan\inst{11}
\and G.C.~Jones \inst{12,13}
\and R.~Amorin\inst{14,15}
\and S.~Bardelli\inst{16}
\and M. Boquien\inst{17}
\and A.~Cimatti\inst{18,19}
\and M.~Dessauges-Zavadsky \inst{1}
\and M.~Giavalisco\inst{20}
\and N.P.~Hathi\inst{21}
\and S. Fujimoto \inst{5,6}
\and E. Ibar \inst{22}
\and A.~Koekemoer\inst{21}
\and G.~Lagache\inst{3}
\and B.C.~Lemaux\inst{23}
\and F. Loiacono\inst{18,19}
\and R.~Maiolino \inst{12,13}
\and D. ~Narayanan\inst{5,24,25}
\and L. Morselli\inst{7,8}
\and Hugo M\'endez-Hern\`andez\inst{22}
\and F. ~Pozzi \inst{18}
\and D. Riechers\inst{26,27}
\and M.~Talia\inst{18,16}
\and S.~Toft\inst{5,6}
\and L.~Vallini\inst{28}
\and D.~Vergani\inst{16}
\and G.~Zamorani \inst{16}
\and E.~Zucca\inst{16}
}
  \institute{Observatoire de Gen\`eve, Universit\'e de Gen\`eve, 51 Ch. des Maillettes, 1290 Versoix, Switzerland
         \and
CNRS, IRAP, 14 Avenue E. Belin, 31400 Toulouse, France
\and
%
%
Aix Marseille Universit\'e, CNRS,  CNES, LAM (Laboratoire d'Astrophysique de Marseille), 13013, Marseille, France
\and
IPAC, California Institute of Technology, 1200 East California Boulevard, Pasadena, CA 91125, USA
\and
Cosmic Dawn Center (DAWN), Copenhagen, Denmark
\and
Niels Bohr Institute, University of Copenhagen, Lyngbyvej 2, DK-2100 Copenhagen, Denmark
\and
Dipartimento di Fisica e Astronomia, Universit\`a di Padova, Vicolo dell’Osservatorio, 3 35122 Padova, Italy
\and
INAF, Osservatorio Astronomico di Padova, vicolo dell’Osservatorio 5, I-35122 Padova, Italy
\and
Kavli Institute for the Physics and Mathematics of the Universe, The University of Tokyo, Kashiwa, Japan 277-8583 (Kavli IPMU, WPI)
\and
Department of Astronomy, School of Science, The University of Tokyo, 7-3-1 Hongo, Bunkyo, Tokyo 113-0033, Japan
\and 
The Caltech Optical Observatories, California Institute of Technology, Pasadena, CA 91125, USA
\and
Cavendish Laboratory, University of Cambridge, 19 J. J. Thomson Ave., Cambridge CB3 0HE, UK
\and
Kavli Institute for Cosmology, University of Cambridge, Madingley Road, Cambridge CB3 0HA, UK
\and
Instituto de Investigacion Multidisciplinar en Ciencia y Tecnologia, Universidad de La Serena, Raul Bitran 1305, La Serena, Chile
\and
Departamento  de Astronomia,  Universidad  de  La
Serena, Av.  Juan Cisternas 1200 Norte, La Serena, Chile
\and
INAF - Osservatorio di Astrofisica e Scienza dello Spazio di Bologna, via Gobetti 93/3, I-40129, Bologna, Italy
\and
Centro de Astronomia (CITEVA), Universidad de Antofagasta, Avenida Angamos 601, Antofagasta, Chile
\and 
University of Bologna, Department of Physics and Astronomy (DIFA), Via Gobetti 93/2, I-40129, Bologna, Italy
\and
INAF - Osservatorio Astrofisico di Arcetri, Largo E. Fermi 5, I-50125, Firenze, Italy
\and
Astronomy Department, University of Massachusetts, Amherst, MA 01003, USA
\and
Space Telescope Science Institute, 3700 San Martin Drive, Baltimore, MD 21218, USA
\and
Instituto de F\'isica y Astronom\'ia, Universidad de Valpara\'iso, Avda. Gran Breta\~na 1111, Valpara\'iso, Chile
\and 
Department  of  Physics,  University  of  California,  Davis,  One  Shields  Ave.,  Davis,  CA  95616,  USA
\and 
Department of Astronomy, University of Florida, 211 Bryant Space Sciences Center, Gainesville, FL 32611 USA
\and
University of Florida Informatics Institute, 432 Newell Drive, CISE Bldg E251, Gainesville, FL 32611
\and 
Department of Astronomy, Cornell University, Space Sciences Building, Ithaca, NY 14853, USA 
\and 
Max-Planck Institut f\"ur Astronomie, K\"onigstuhl 17, D-69117, Heidelberg, Germany
\and
Leiden Observatory, Leiden University, PO Box 9500, 2300 RA Leiden, The Netherlands
\\
\email{daniel.schaerer@unige.ch}
}

\authorrunning{D.\ Schaerer et al.}
\titlerunning{ALPINE \cii--SFR relation at high redshift}

\date{Received 29 january 2020; accepted 29 april 20202}

\abstract{The \Cii\ line is one of the strongest IR emission lines, which has been shown to trace the star formation rate (SFR) of 
galaxies in the nearby Universe, and up to $z \sim 2$. Whether this is also the case at higher redshift and in the early Universe remains debated.
The ALPINE survey, which targeted 118  star-forming galaxies  at $4.4 < z< 5.9$, provides a new opportunity to examine this question with the first statistical dataset. 
Using the ALPINE data and earlier measurements from the literature, we examine the relation between the \cii\ luminosity and the SFR
over the entire redshift range from $z \sim 4-8$.
ALPINE galaxies, which are both detected in \cii\ and in dust continuum, show good agreement with the local \lcii--SFR relation.
Galaxies undetected in the continuum by ALMA are found to be over-luminous in \cii when the UV SFR is used.
After accounting for dust-obscured star formation, by an amount of SFR(IR)$\approx$SFR(UV) on average, which results from two different 
stacking methods and SED fitting, the ALPINE galaxies show an \lcii--SFR relation comparable to the local one.
When \cii\ non-detections are taken into account, the slope may be marginally steeper at high-$z$, although this is still
somewhat uncertain. 
When compared homogeneously, the $z>6 $ \cii\ measurements (detections and upper limits) do not behave
very differently to the $z \sim 4-6$ data.
We find a weak dependence of \lcii/SFR on the \lya\ equivalent width. 
Finally, we find that the ratio \lcii /\lir $\sim (1-3) \times 10^{-3}$  for the ALPINE sources, comparable to that of `normal' galaxies at lower redshift.
Our analysis, which includes the largest sample ($\sim 150$ galaxies) of \cii\ measurements at $z>4$ available so far,
suggests no or little evolution of the \cii--SFR relation over the last 13 Gyr of cosmic time.
}

 
\maketitle

\section{Introduction}
\label{s_intro}
The \Cii\ line is an important coolant of the neutral interstellar medium (ISM), one of the strongest
emission lines in the infrared (IR), which is also emitted relatively close to the peak of dust continuum
emission. Although \cii\ has long been known to originate from \hii\ regions, diffuse neutral and ionised ISM, and
from photodissociation regions \citep[e.g.][]{Wolfire1995The-Neutral-Ato,Hollenbach1999Photodissociati},
it has been found to trace star formation. In particular, the \cii\ luminosity has been shown to correlate well with the total star formation 
rate (SFR) of galaxies in our Galaxy, nearby galaxies, and up to $z \sim 2$ \citep[see e.g.][and references therein]{Pineda2014A-Herschel-C-II,Herrera-Camus2015C-II-158-mum-Em,2011MNRAS.416.2712D,De-Looze2014The-applicabili}.
Other studies have recently stressed that \cii\ could alternatively be used as a tracer of molecular gas
\citep[see e.g.][]{Nordon2016C/H2-gas-in-sta,Glover2016CO-dark-gas-and,Fahrion2017Disentangling-t,Zanella2018The-C-ii-emissi}.

Since \Cii\ can be observed from the cosmic noon ($z \sim 2$) out to very high redshift \citep[$z \sim 7-8$][]{Inoue2016Detection-of-an} with ALMA, 
and potentially even into the cosmic dark ages with other facilities \citep[cf.][]{Carilli2017Galaxies-into-t},
this line has often been targeted. The goal is to use it as a probe of the ISM properties in distant galaxies,
as a measure of the total SFR (unaffected by the possible presence of dust), and for other purposes, including redshift confirmation 
for galaxies in the epoch of reionisation.

The first attempts to measure \Cii\ in galaxies at $z>6$ with ALMA have mostly been unsuccessful,  essentially yielding
non-detections, both for \lya\ emitters (LAEs) and Lyman-break galaxies (LBGs) \citep[e.g.][]{Ouchi2013An-Intensely-St,Ota2014ALMA-Observatio,Maiolino2015The-assembly-of}.
Subsequent observations have detected \cii\ in LAEs and LBGs, both in blank fields and behind lensing clusters,
finding several \cii-underluminous galaxies at high-$z$ and suggesting a large scatter in \lcii-SFR
\citep[see e.g.][]{Maiolino2015The-assembly-of,Willott2015Star-Formation-,Pentericci2016Tracing-the-Rei,Bradac2017ALMA-C-II-158-m,Carniani2018Kiloparsec-scal} compared to the local samples \citep{De-Looze2014The-applicabili}.
On the other hand, \cite{Riechers2014ALMA-Imaging-of} and \cite{Capak2015Galaxies-at-red} successfully detected several $z \sim 5-6$ 
star-forming galaxies, revealing relatively broad \cii\ lines and a good agreement with the local \cii--SFR relation.
Reanalysing the existing \cii\ detections and non-detections of $z \sim 6-7$ galaxies, \cite{Matthee2019Resolved-UV-and}
showed that the available data appears compatible with the \citedelooze\ relation for SFR$\ga 30$ \msunyr\
and may deviate from that of lower SFRs, if broader \cii\ lines are assumed for the non-detections and the data are consistently compared.
Conversely, using very similar data, \cite{Harikane2018SILVERRUSH.-V.-} and \cite{Harikane2019Large-Populatio}
concluded that $z=5-9$ galaxies show a clear \cii\  deficit with respect to the local \cii--SFR relation, and that this deficit 
increases with increasing \lya\ equivalent width.
Manifestly, no consensus has yet been reached on these questions, and it is unclear if the \Cii\ line remains a good tracer
of star formation at $z> 4$ or if there is a quantitative change compared to the observations at low redshift.

To make progress on these issues, we  used the ALMA Large Program to INvestigate $C^+$ at Early Times (ALPINE) survey, 
which targets 118 `normal' (i.e.\ main sequence) star-forming galaxies with known spectroscopic redshifts at $4.4 < z< 5.9$, and 
which is designed to provide the first statistical dataset making it possible to determine the observational properties
of \cii\ emission at high-$z$. The survey was recently completed and is described in detail in
\citesurvey, \citedata, and \citeancillary.
Our measurements, yielding 75 high-significance detections of \Cii\ and 43 non-detections, combined with
the earlier \cii\ observations of 36 galaxies at $z \sim 6-9.11$ from literature compilations, allow us to
examine what is normal for high-$z$ galaxies and shed new light on the above questions.

The paper is structured as follows. 
We briefly summarise the ALPINE \cii\ dataset and other measurements in Sect.\ \ref{s_obs}.
We then examine the behaviour of the \Cii\ luminosity with different SFR indicators, and we
carefully compare the \cii--SFR observations of high-$z$ galaxies to the reference sample of 
\citedelooze\ (Sect.\ \ref{s_cii}). We combine the ALPINE dataset with the available
\cii\ observations at $z>6$ and examine whether all high-redshift observations show the same
picture and if the \cii--SFR relation is different in the early Universe (Sect.\ \ref{s_universal}).
Finally, we present the observed \cii-to-IR ratio in Sect.\ \ref{s_cii_ir}. 
We discuss the possible caveats and future improvements in Sect.\ \ref{s_discuss}.
Our main results are summarised in Sect.\ \ref{s_conclude}, and we provide results to fits to different
datasets in the Appendix.
We assume a $\Lambda$CDM cosmology with $\Omega_\Lambda = 0.7$, $\Omega_m=0.3$ and $H_0 = 70$ \kmsmpc,
and a Chabrier IMF \citep{Chabrier2003Galactic-Stella}.

\section{Observations and derived quantities}
\label{s_obs}

The ALMA Large Program to INvestigate \cii\ at Early times (ALPINE) survey, presented  in  \citesurvey, has observed 118 `normal' 
star-forming galaxies with known spectroscopic redshifts at $4.4 < z< 5.9$. 
The ALPINE sample also includes seven galaxies (HZ1, HZ2, HZ3, HZ4, HZ5, HZ6/LBG-1, and HZ8) that were previously 
observed with ALMA by \cite{Riechers2014ALMA-Imaging-of} and \cite{Capak2015Galaxies-at-red}.
It currently constitues the largest sample of \cii\ observations at $z \sim 4-6$.

Details of the ALPINE data reduction and statistical source properties are described  in
\citedata, from which we used the \Cii\ line luminosities ($L(\cii)$, 75 detections with high significance, and 43 non-detections)
and the dust continuum measurements  (23 detections and 95 non-detections). The 158 \micron\ rest-frame continuum fluxes have been 
converted to total IR luminosities, \lir, using  an average empirically based conversion from the 158 \micron\ monochromatic continuum 
flux density to \lir\  as described in \citedata. 
The empirical template gives a conversion similar to  a modified black body with a dust temperature of $T_d = 42$ K, 
a dust opacity at 850 $\mu$m of $k_{850} = 0.077 ~ {\rm m^2 ~kg^{-1}}$, and a grey-body power-law exponent $\beta = 1.5$ \citep[see e.g.][]{Ota2014ALMA-Observatio,Matthee2019Resolved-UV-and}.

For galaxies undetected in \cii,\ we used the `aggressive' $3 \sigma$ upper limits of \lcii\ reported in \citedata, 
defined as three times the RMS of the noise in velocity-integrated flux maps obtained by collapsing a channel width of 300 \kms\ centered around the expected spectroscopic redshift.
We then rescaled these limits to reflect a more realistic (though less conservative) distribution of full width half maximum (FWHM)
of our \cii-undetected galaxies: motivated by the observed dependence of FWHM on \lcii\ shown in Fig.\ 1
\footnote{A linear fit to the data yields $\log(\lcii/\lsun)=2.24 \times \log({\rm FWHM/(km s^{-1})} + 3.21$.}, we adopted FWHM $= 150$ \kms, instead of the median of $252$ \kms\ of the \cii-detected ALPINE galaxies (\citedata).
As discussed in \citedata, we note that by construction our 3 $\sigma$ upper limits of \lcii\ can be underestimated if the sources are: (i) just below the detection threshold; (ii) spatially extended (larger than the beam-size, e.g., $\gtrsim 1''$); and/or (iii) show very broad line profiles (see e.g. Kohandel et al. 2019), where the two latter conditions are less likely to occur in less massive and less star-forming objects.

For galaxies undetected in continuum, we used the aggressive upper limits determined by \citedata, using the same 
conversion from158 \micron\ rest-frame continuum fluxes to total IR luminosity.
Finally, we also used \lir\ values derived from the IRX--$\beta$ relation obtained by stacking 
of the ALPINE sources, as described in \cite{Fudamoto2020The-ALPINE-ALMA}. 
In the absence of a direct detection of dust continuum emission, this is our preferred method to correct for dust-obscured star formation.

\setcounter{figure}{0}
\begin{figure}[tb]
{\centering
\includegraphics[width=9cm]{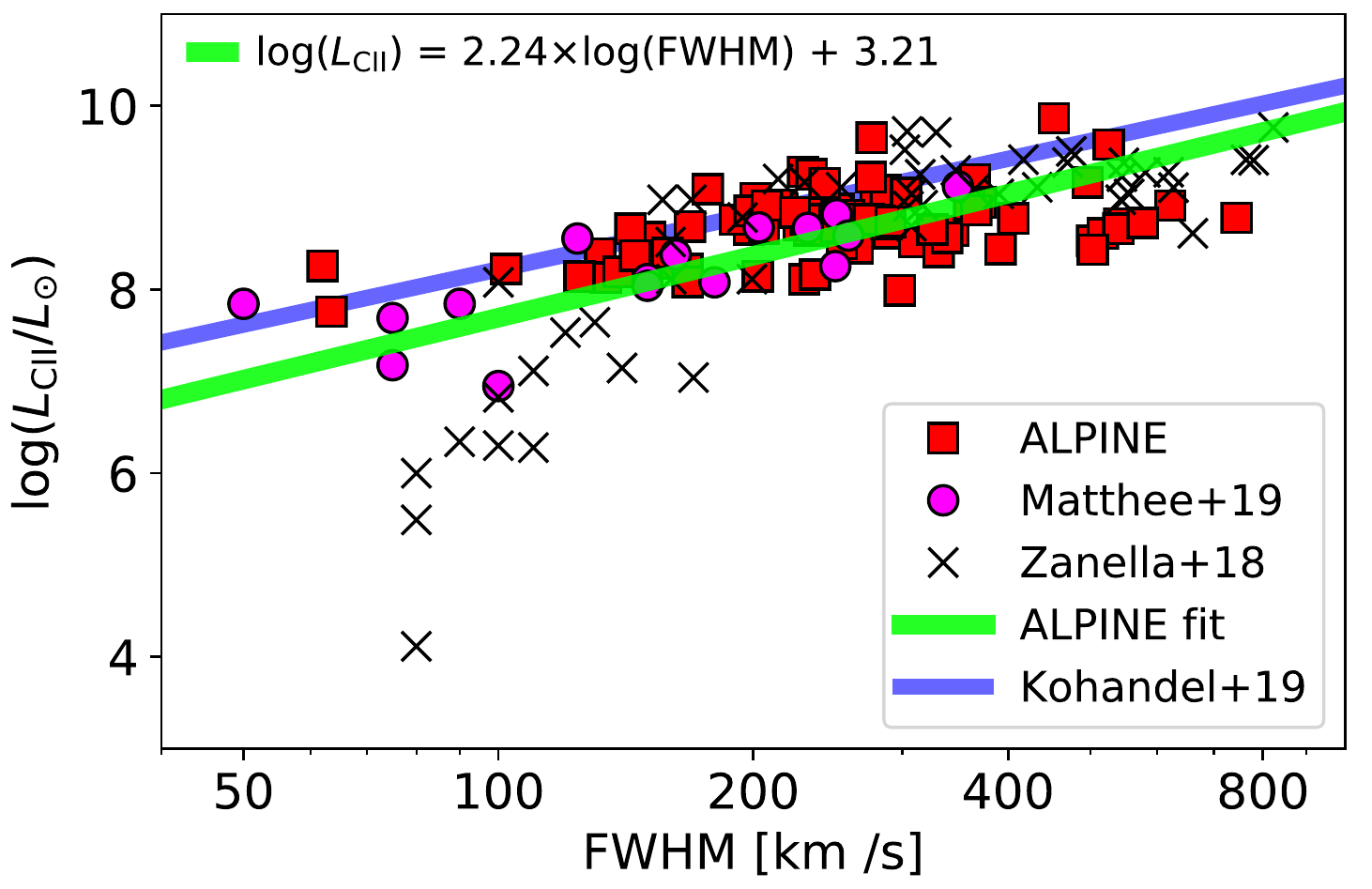}
\caption{Observed FWHM of the \Cii\ line as function of the \cii\ luminosity from ALPINE and 
other data from the literature (the compilation of $z>6$ sources from \cite{Matthee2019Resolved-UV-and}
and available measurements for galaxies in the \cite{Zanella2018The-C-ii-emissi} compilation ($z \sim 0-6$).
The violet dashed line is a  Tully-Fisher-like relation derived for high-$z$ galaxies by \cite{Kohandel2019Kinematics-of-z}.
The green solid line shows our best fit to the data.
}}
\label{fig_fwhm}
\end{figure}

From the rich dataset of ancillary photometric and spectroscopic data, which is also available for the ALPINE sources
(see \citeancillary\ for details), we used the observed UV luminosity (or equivalently the absolute UV magnitude \muv\ 
at 1500 \AA). 
To compare our \cii\ data with other measurements and results in the literature, we also used measurements
of the \lya\ equivalent widths, \ewlya, obtained from the rest-UV spectra of our sources, which were obtained during earlier
spectroscopic observations with DEIMOS and VIMOS on the Keck and VLT telescopes. The spectra are
discussed by \citeancillary. The \lya\ measurements, available for 98 sources, are taken from \citecassata, where
a more detailed description of the \lya\ properties is presented.

 From the above-mentioned measurements of the UV and IR luminosities, we derived three
 `classical' measures of star-formation rate, SFR(UV) uncorrected for attenuation, SFR(IR),
 and the total SFR(tot)$=$SFR(UV)$+$SFR(IR). 
 We also used estimates of the total SFR, SFR(SED), obtained from the multi-band SED fits of \citeancillary.
 The different SFR measurements are all included in the ALPINE database\footnote{ \url{https://cesam.lam.fr/a2c2s/}},
 where the data will be made public.
 
 To allow a proper comparison of the \lcii--SFR relation with the low-$z$ galaxy sample of \citedelooze, 
 we adopted the same conversion factors between \luv, \lir, and SFR as the ones used in their paper. 
 We note that the SFR(UV) calibration adopted by \citedelooze\  agrees with the classical one from \cite{kennicutt1998}, when rescaled to the same IMF. However, for the same IMF, their IR calibration, taken from \cite{Murphy2011Calibrating-Ext}, yields 30\% (0.12 dex) higher SFR(IR) values  than the  
\cite{kennicutt1998} calibration.
Finally, we rescaled the SFR(UV) and SFR(IR) values by a factor of 1.06 from the Kroupa IMF (used by \citedelooze) to the Chabrier IMF, 
for consistency with the other ALPINE papers.\footnote{In short, the final adopted SFR calibrations are: 
SFR(UV)/(\msunyr)$=8.24 \times 10^{-29} L_\nu$, where $L_\nu$ is in units of ergs/s/Hz, 
or equivalently SFR(UV)/(\msunyr)$=1.59 \times 10^{-10} \luv/\lsun$, where \luv\ is calculated at 1500 \AA. 
And SFR(IR)/(\msunyr)$=1.40 \times 10^{-10} \lir/\lsun$.} 
We highlight that we assume SFR(IR)=0 per default and unless otherwise stated for sources that are not detected in the continuum. 
This is discussed further below.

\begin{figure}[tb]
{\centering
\includegraphics[width=9cm]{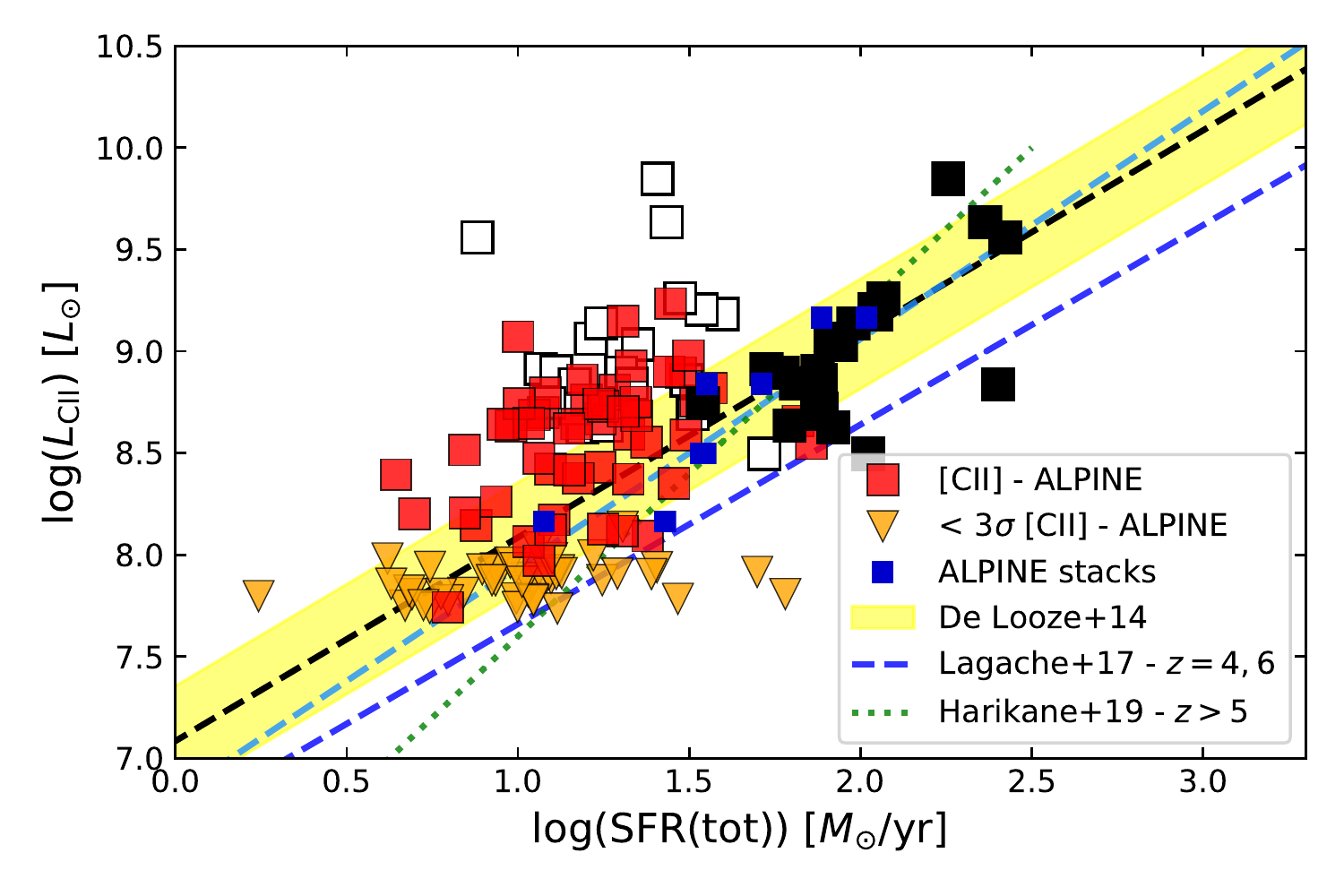}
\caption{\cii\ as a function of UV or UV+IR-derived SFR for the $z \sim 4.5$ ALPINE sources. 
Squares show the \cii\ detections, orange triangles the 3 $\sigma$ upper limits.
Black squares show galaxies with continuum detection (black filled squares show SFR(UV)+SFR(IR), empty squares the SFR(UV) of the same sources); 
red squares shows the SFR(UV) for the other (non-continuum-detected) ALPINE sources.
Blue circles show the results from stacks of ALPINE sources in four bins of $L(\cii)$ and two redshift bins, adapted from 
from \citedata.
The observations are compared to the \cii\--SFR relations of local galaxies determined by
\cite{De-Looze2014The-applicabili} adjusted to the Chabrier IMF by reducing the SFR by a factor of 1.06 (black dashed line), 
shown by the yellow band with a total width corresponding to $2 \sigma$.
The green dotted line shows the relation fitted to observations of $z \sim 5 - 9$ galaxies by \cite{Harikane2019Large-Populatio}.
The fits from the models of \cite{Lagache2018The-CII-158-mum} for 
redshifts spanning the range of the observations are shown by the two blue dashed lines.
}}
\label{fig_cii_sfr}
\end{figure}

\begin{figure*}[tb]
{\centering
\includegraphics[width=9cm]{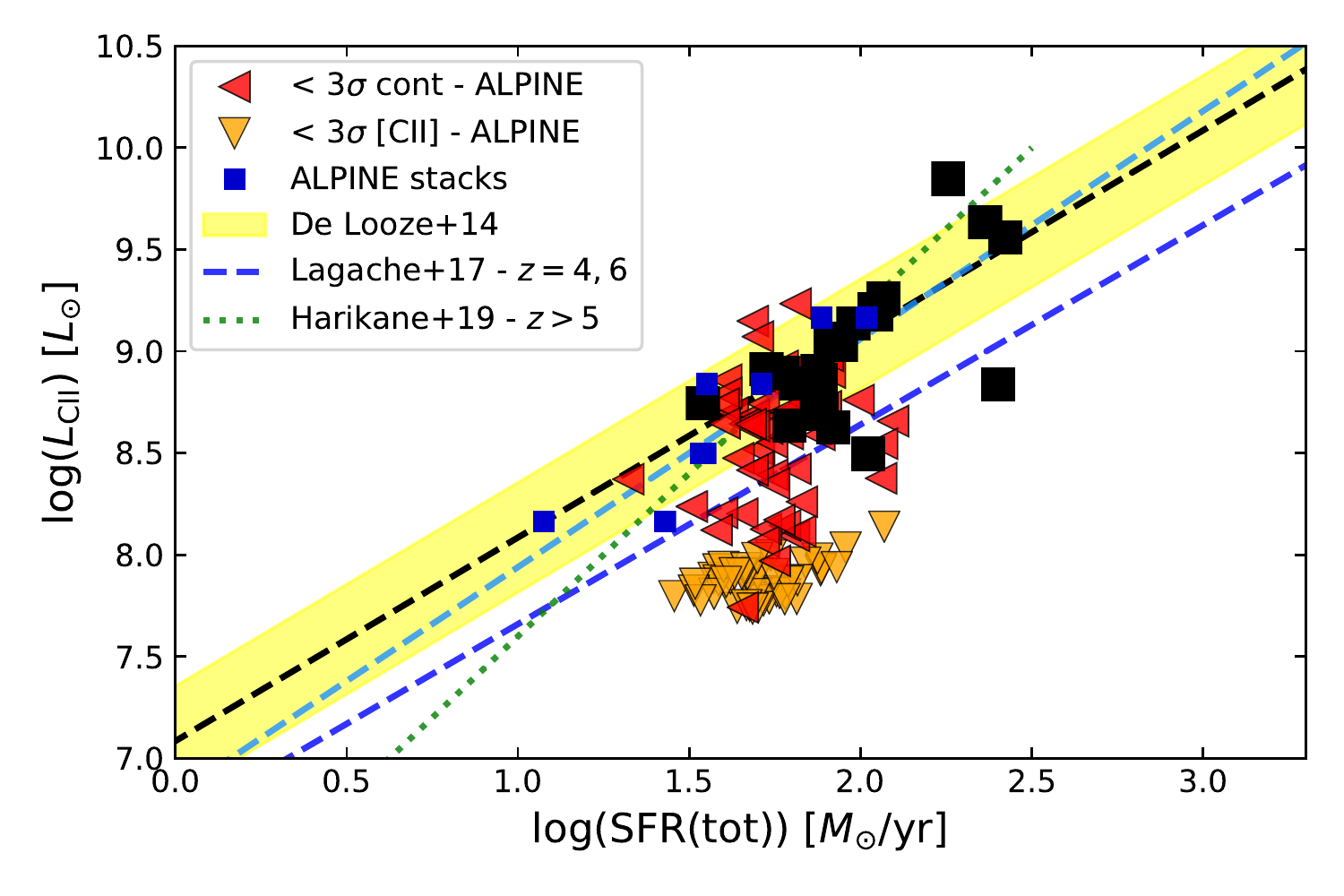}
\includegraphics[width=9cm]{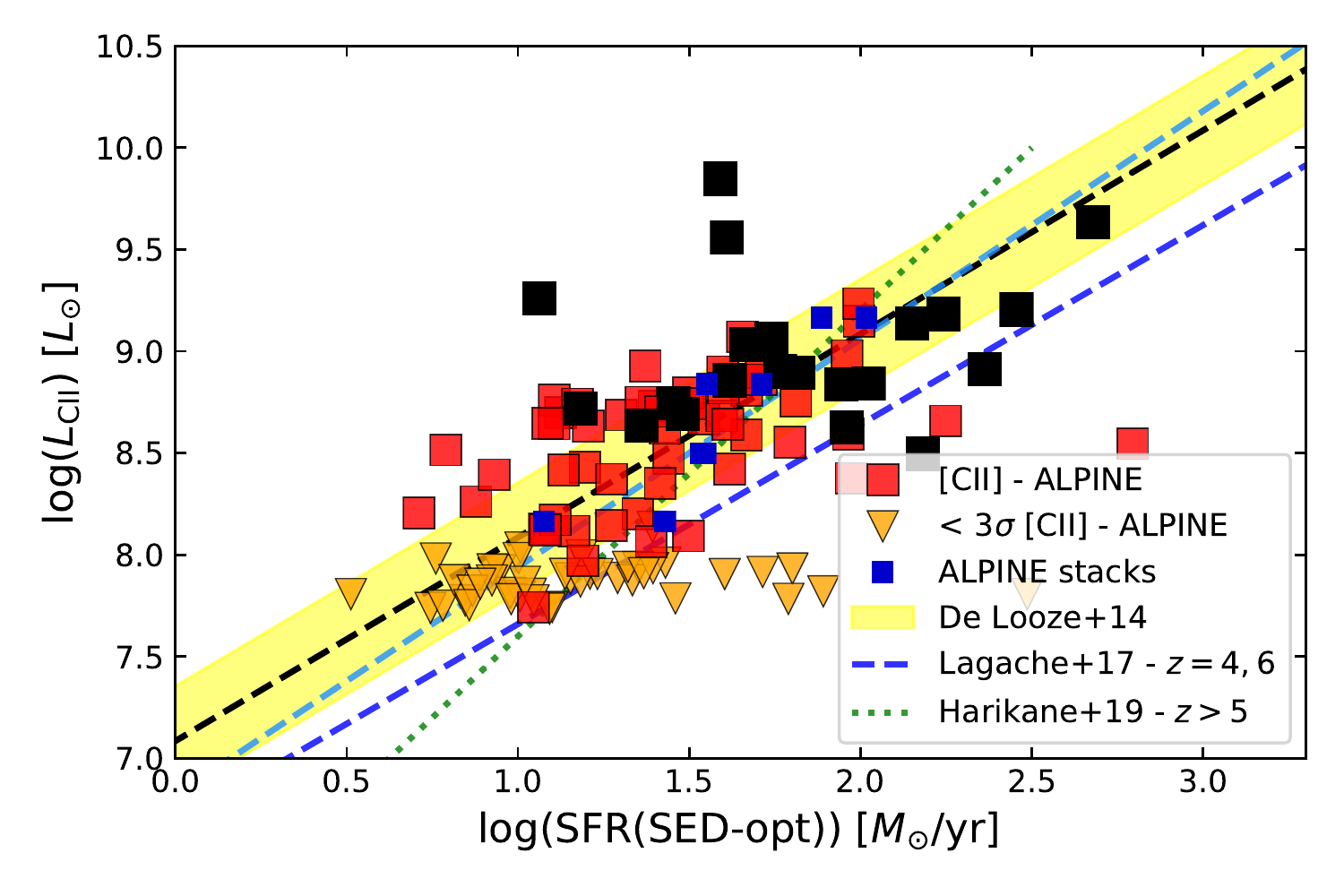}
\caption{Same as Fig.\  \protect\ref{fig_cii_sfr}. {\em Left:} SFR(UV+IR) where the IR contribution now includes the 1-$\sigma$ limit
on \lir. {\em Right:} using the SFR derived from SED fitting of the stellar emission (rest-UV to optical).}}
\label{fig_3}
\end{figure*}

\begin{figure}[tb]
{\centering
\includegraphics[width=9cm]{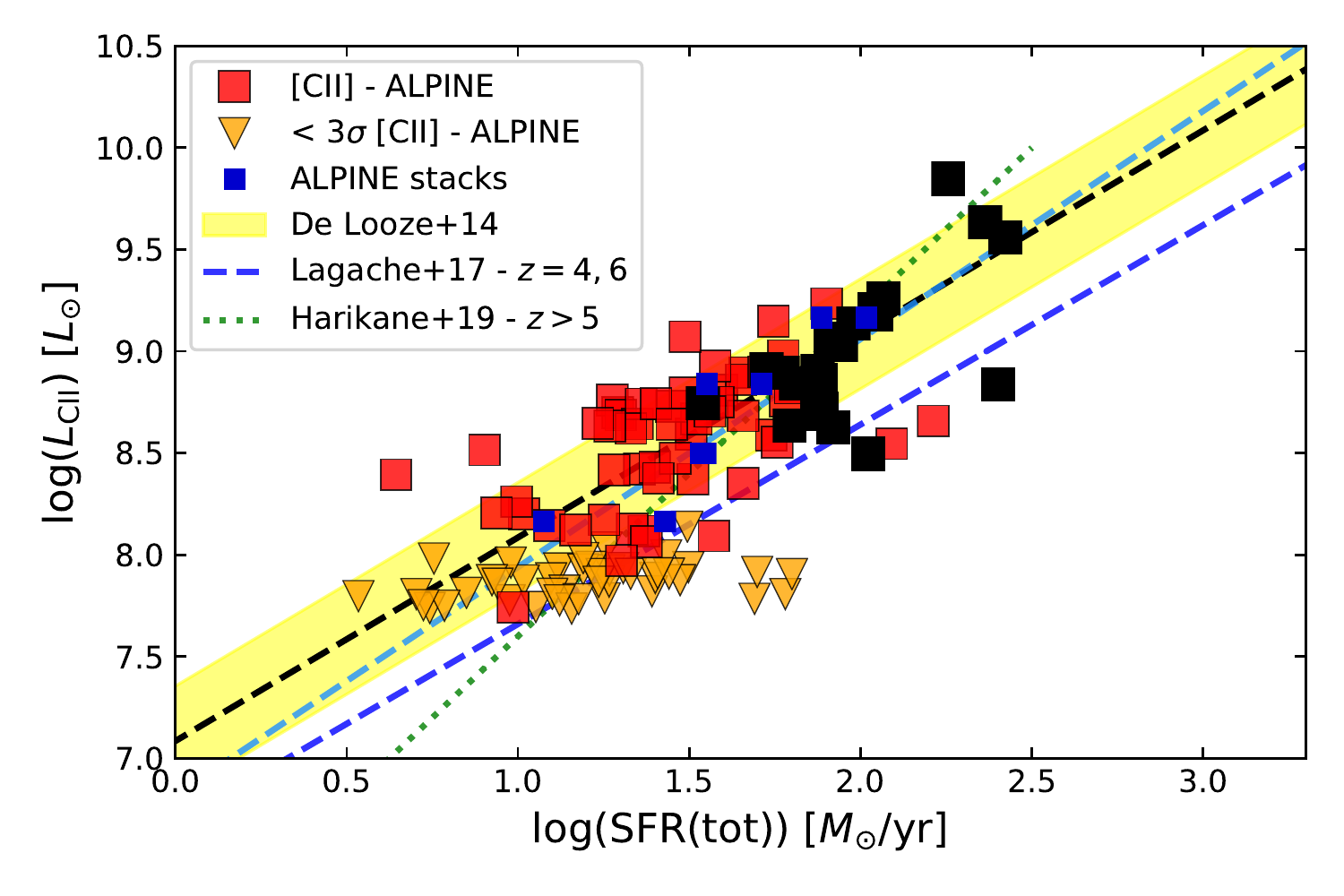}
\caption{Same as Fig.\  \protect\ref{fig_cii_sfr} using SFR(UV+IR). Here, SFR(IR) is derived from the observed UV slope and luminosity using 
the ALPINE IRX-$\beta$ relation obtained from stacking \citep{Fudamoto2020The-ALPINE-ALMA}. The \cii-\-detected galaxies follow the local relation well.}}
\label{fig_4}
\end{figure}

\section{Relations between the \Cii\ luminosity and SFR indicators at $z \sim 4-6$ and higher redshift}
\label{s_cii}

\subsection{Comparing \lcii\ with UV, IR, and SED-fit-based SFRs}

As is often done for high-$z$ galaxies, which are generally selected from the rest-UV and seldomly detected 
in the dust continuum, we first used a basic SFR indicator, SFR(UV) derived from the observed UV luminosity, which 
is available for the entire sample, to obtain the \lcii--SFR relation shown in Fig.\ \ref{fig_cii_sfr}.
The ALPINE data is compared to average relation of the low-$z$ \hii-galaxy/starburst sample from \citedelooze\ as a reference (henceforth 
named the `local' relation), which is often used in the literature.
It includes 184 galaxies, shows a linear scaling between \lcii\ and SFR, and a
scatter of 0.27 dex (see their Table 3).\footnote{For comparison, their entire sample of  530 galaxies shows  a larger
scatter (0.42 dex) and a \Cii\ luminosity, which is lower by 0.07 dex for a given SFR.}
While the \cii\ detections span a wide range between $\lcii \sim 5 \times 10^7$ \lsun\ and $5 \times 10^9$ \lsun,
SFR(UV) varies less, thus resulting in a relatively  steep relation between \lcii\ and SFR(UV).
Compared to the local \lcii--SFR correlation, the \cii\ luminosity of our sources appears higher,
in contrast to several high-$z$ ($z \ga 6$) galaxies where \cii\ was found to be `under-luminous',
as mentioned in the introduction.
It is more likely that  the SFR is underestimated, as can be expected from dust attenuation of the UV light.

To correct for dust attenuation in the simplest way, in the same figure (Fig.\ \ref{fig_cii_sfr})
we plot the \cii\ measurements as a function 
of the total SFR, adding the dust-attenuated SFR(IR) to SFR(UV) for the galaxies for which
we detect emission from the dust continuum. 
Clearly, for the continuum-detected sources, the increase in SFR is significant, bringing them
to fair agreement with the local \lcii--SFR relation, as seen by the comparison with the left panel.
This corresponds to galaxies with SFR(tot) $\ga 30$ \msunyr.

On average, however, the \cii\ luminosities of the 74 detected sources remain larger than expected from the 
local relation of \citedelooze,  by a factor $\sim 1.5$ for the entire sample and
a factor $\sim 2$ for the sources that are not detected in the continuum (red squares in 
Fig.\ \ref{fig_cii_sfr}). 
Approximately 40\% of the \cii-detected ALPINE galaxies are extended and classified as
mergers from a morphological and kinematic analysis (\citesurvey). 
Excluding, for example, these mergers from the sample does not significantly change the 
deviation from the relation; on average, a shift by a factor 1.25 in SFR(tot) remains, compared 
to a factor 1.5 shift for the entire sample.
For the mergers alone, the deviation is 0.28 dex, similar to that of several sources not detected in the continuum.
From this, we conclude that even if there were systematic differences between mergers and galaxies in 
the local sample, this would probably not explain the observed deviation between the ALPINE dataset
and the \citedelooze\ relation.
Obscured star formation, below our current detection threshold in the ALMA measurements, is probably 
present in the majority of the ALPINE sample.

\citedata\ carried out stacking of the ALPINE sources in different bins of \cii\ luminosity, detecting thus the dust continuum
in several of these bins, and hence measuring in particular the average dust-attenuated contribution SFR(IR). 
After conversion to the same SFR calibrations used here (cf.\ above)  their results are shown in Fig.\ \ref{fig_cii_sfr}.
The ALPINE stacking results show a good agreement with the local \cii--SFR relation, indicating that some correction 
for dust-obscured star formation is necessary even for the continuum undetected galaxies, and especially those at the low \lcii\ range.

Regrettably, the upper limits on the IR continuum fluxes of the individual ALPINE sources are not sufficiently constraining.
Even if we use the aggressive $1 \sigma$ limits to determine
a limit on hidden star formation by summing SFR(UV) and the SFR(IR) limit,  we obtain the SFR(tot) limits
shown in Fig.\ 3  
(left), which are mostly in the range of SFR(tot)$\la 40-100$ \msunyr. 
Clearly, tighter constraints are desirable to examine if/how the high-$z$ galaxies do or do not deviate from the local
\cii--SFR relation.
Fitting the SED provides one way to account for this. Using tihe results from the multi-band SED fitting
results discussed in \citeancillary\ and assuming a Calzetti attenuation law, reduces the apparent
\cii\ excess found when not accounting for dust-obscured star formation. As shown in Fig.\ 3
(right), the mean offset between the local \lcii--SFR relation and the ALPINE data
(\cii\ detections only) is then reduced to 0.06 dex, although the scatter around the local relation is quite large (0.40 dex).
However, we note that a comparison using SED-based SFR values with the relations established
by \citedelooze\ would be methodologically inconsistent, since these authors use simple (UV and IR) SFR calibrations,
whereas SED fitting allows for varying star formation histories, different ages, etc., which may
not yield compatible results and is known to give a larger scatter \citep[see e.g.][]{wuytsetal2011,schaerer2013}.
In any case, all the methods illustrated here show that most, if not all of the $z \sim 4-6$ galaxies included
in the ALPINE sample must suffer from some dust attenuation.

\subsection{The \cii--SFR relation accounting for hidden SF in $z \sim 4-6$ galaxies}
\label{s_hidden}

We now proceed to account for hidden SF in all individual ALPINE galaxies in the best possible 
and consistent way to compare the \cii--SFR relation with lower redshift data.
To do this, we used the average IRX--$\beta$ relation derived by \cite{Fudamoto2020The-ALPINE-ALMA} for the ALPINE sample from median 
stacking of the continuum images in bins of the UV slope $\beta$.
These authors found an IRX--$\beta$ relation, which is close to but below the relation expected 
for the SMC attenuation law, with little evolution across the redshift range of the ALPINE sample.
We applied their mean IRX--$\beta$ relations to each individual source for which the
dust continuum had not been detected, thus yielding a predicted \lir\footnote{We note that the  \lir\ values 
predicted in this way are below the conservative upper limits determined from the individual non-detections.}
and hence a corresponding SFR(IR), using the same assumptions as for the rest of the sample. 
For continuum-detected galaxies, we used the standard SFR(IR) values, shown above.

The result is illustrated in Fig.\ 4
showing a very small offset from 
the local relation ($-0.05$ dex) and a scatter of 0.28 dex around it (for the \cii\ detections).
In other words, taking into account a relatively small correction for hidden SF, which is compatible with
our continuum non-detections and the IRX--$\beta$ relation, the ALPINE \cii-detected galaxies nicely follow
the same \lcii--SFR relation as low-redshift galaxies.

Now, if we include the `agressive' \cii\ non-detections and fit the data with a linear relation of the form
\begin{equation}
\log(\lcii/\lsun) = a + b \times \log({\rm SFR}/\msunyr),
\end{equation}
using a Bayesian fit including censored data\footnote{We follow the method of \cite{Kelly2007Some-Aspects-of} implemented 
in the python package {\em linmix}, \url{https://github.com/jmeyers314/linmix}.
The method allows for uncertainties in one quantity, here \lcii.},
we obtain a slope that is marginally steeper than the local relation ($1.17 \pm 0.12$, the results from different fits are given in the Appendix).
However, adopting more conservative upper limits for \lcii, for example\ the `secure' limits\footnote{Secure limits are calculated by summing the 3 $\sigma$ rms of the noise to the highest flux measured in 1 arcsec around the phase centre in visibility-tapered velocity-integrated flux maps (see \citedata\ for more details).} from \citedata, which are typically
less deep by a factor of two, our fits including censored data yields a slope of $0.96 \pm 0.09$, compatible with unity, 
and slightly offset (by $\sim - 0.03$ dex) with respect to the local relation.
From this we conclude that main sequence galaxies at $z \sim 4-6$ may show the same \lcii--SFR relation
as low redshift galaxies or a relation which is somewhat steeper (with an exponent $\sim 1.2$). 
More firm statements are difficult to make at the present stage, until \cii\ non-detections and the exact amount 
of dust-obscured star formation are better quantified.

\begin{figure}[tb]
{\centering
\includegraphics[width=9cm]{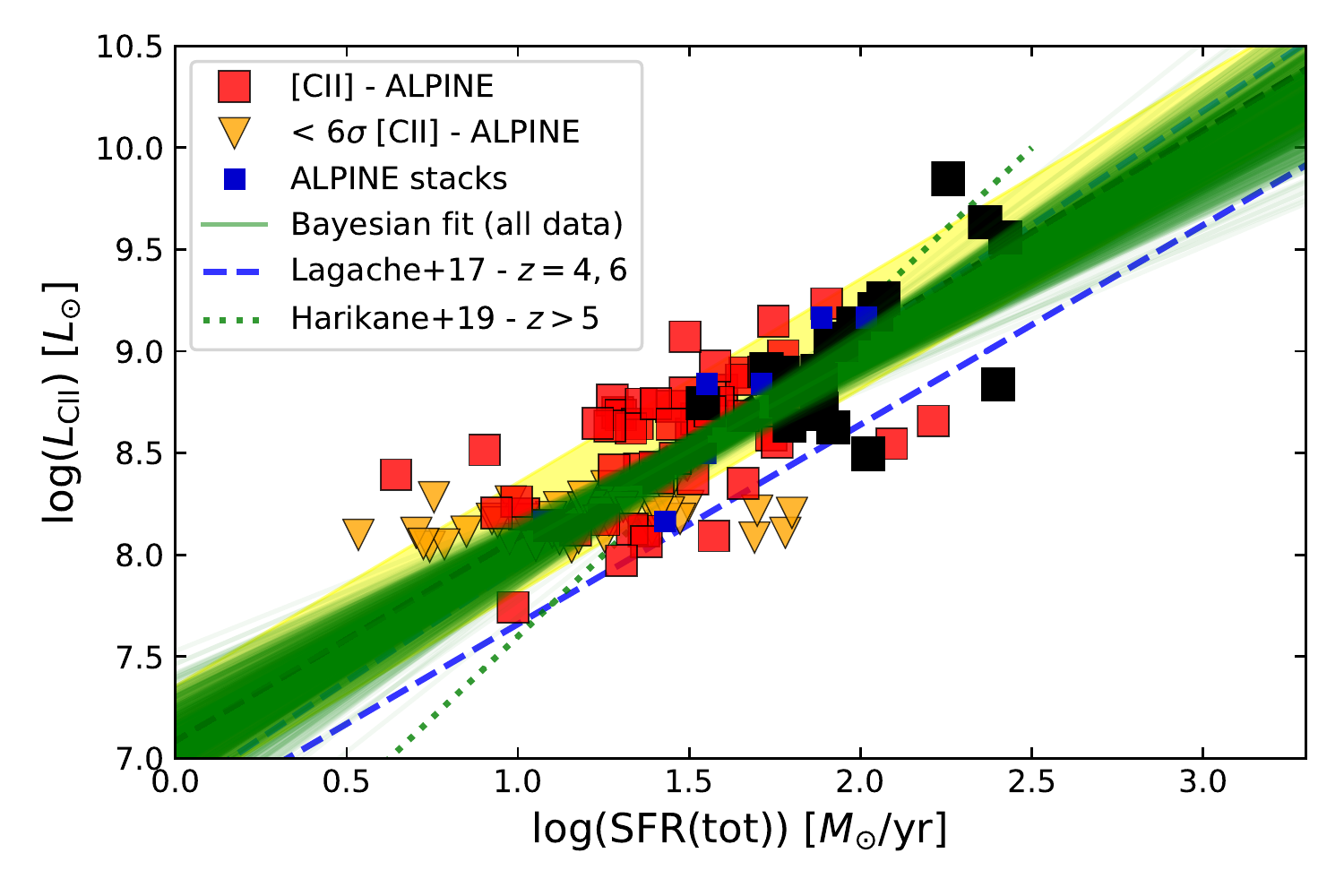}
\caption{Same as Fig.\ 4
 for ALPINE sources, but adopting conservative upper limits for the \cii\ non-detections 
(two times the agressive $3 \sigma$ upper limits).
The Bayesian linear fit to all the measurements (detections and upper limits) is shown by the dark green lines/band,
which also illustrates the probability distribution of the fit. The fit yields a slope of  $0.96 \pm 0.09$, compatible with unity, 
and a small but insignificant offset (by $\sim -0.03$ dex) with respect to the local relation.
}}
\label{fig_cii_conservative}
\end{figure}

\begin{figure}[tb]
{\centering
\includegraphics[width=9cm]{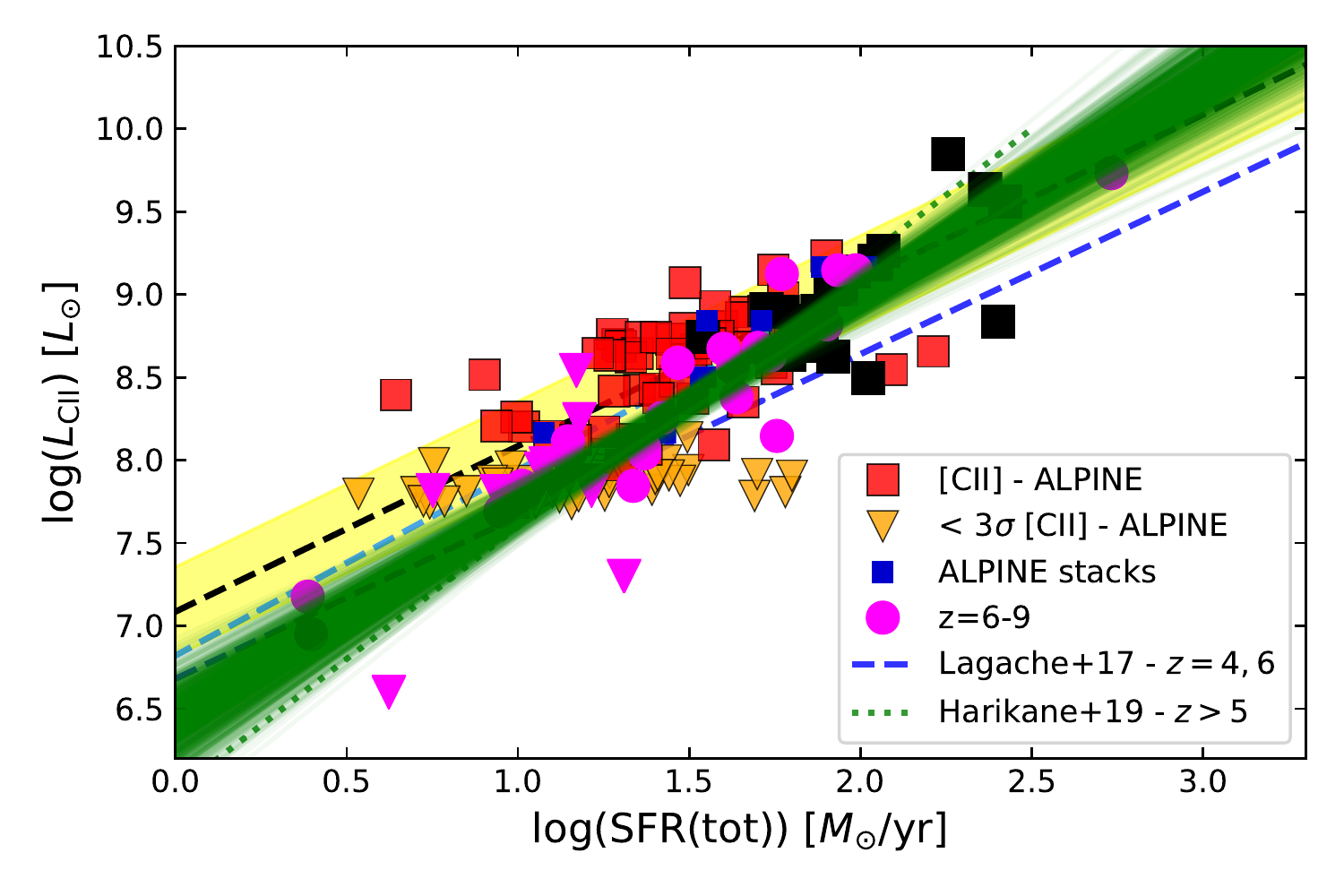}
\caption{\cii--SFR relation combining the  ALPINE sample and observations of $z \sim 6-9$ galaxies 
taken from the literature. The $z >6$ data, after proper uniformisation, are plotted with pink symbols (circles for detections, and triangles for upper limits).
The SFR of all sources includes an SFR(IR) contribution, determined from the observed
dust continuum or from the IRX--$\beta$ relation if undetected. 
All \cii\ non-detections are illustrated by triangles, showing $3 \sigma$ (aggressive) upper limits.
The Bayesian linear fit to all the $z>4$ measurements (detections and upper limits) is 
shown by the dark green lines/band. The slope is somewhat steeper than unity: $1.28 \pm 0.10$.}}
\label{fig_cii_z69}
\end{figure}

\begin{figure}[tb]
\includegraphics[width=9cm]{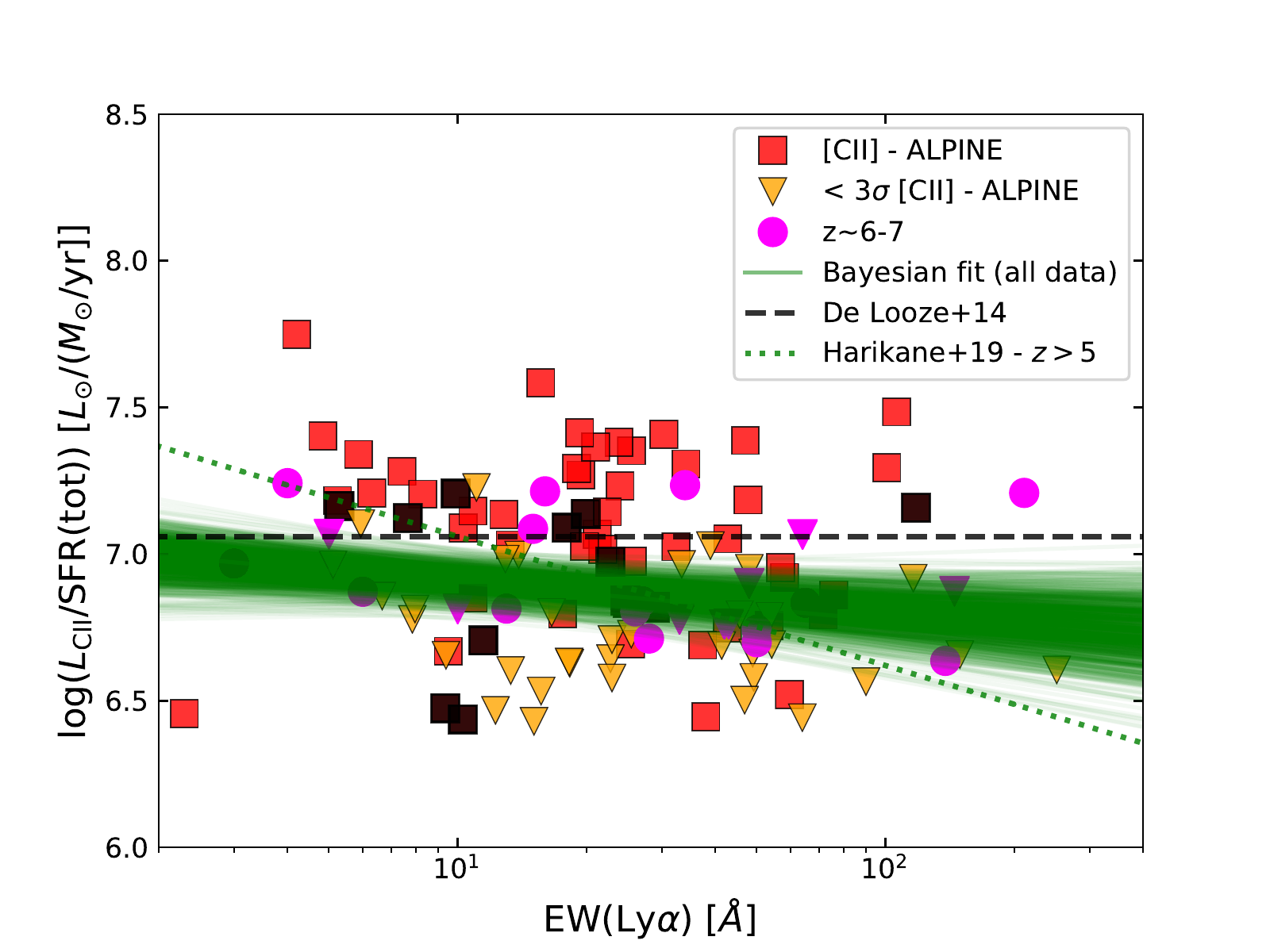}
\caption{$L(\cii)/{\rm SFR(tot)}$ as a function of the rest-frame \lya\ equivalent width of the ALPINE
sources and  $z \sim 6-7$ galaxies taken from the literature (pink symbols: circles for detections, and triangles for upper limits).
The dotted line shows the fitting relation obtained by \cite{Harikane2018SILVERRUSH.-V.-}
for a compilation of $z \sim 5.7-7$ galaxies, the horizontal line the average value from the local relation.
The green lines/band show the fit to the data; the resulting slope is $-0.11 \pm 0.06$, indicating 
a weak dependence on EW(\lya).}
\label{fig_cii_lya}
\end{figure}

\subsection{Is there a universal behaviour of \lcii\ at $z>4$?}
\label{s_universal}

We now examine how the ALPINE \cii\ measurements of $z \sim 4-6$ galaxies compare
with the other available observations at even higher redshifts.
To do this, we use the recent compilation of \cite{Matthee2019Resolved-UV-and}, which 
includes 25 reported ALMA \cii\ observations of galaxies with known spectroscopic redshifts between
$z=6.0$ and $z=7.212$.  Importantly,  \cite{Matthee2019Resolved-UV-and} recomputed
\cii\ non-detection limits using empirically-motivated \cii\ line widths\footnote{They use FWHM$=-1215-66 \times M_{\rm UV}$ \kms,  
translating to a minimum FWHM$=123$ \kms\ for the UV-faintest \cii\ non-detected galaxy in their sample 
(Matthee, private communication).}. Furthermore they have uniformly re-derived SFR(UV) and SFR(IR) from 
the observations, assuming a modified black body with $T_d=45$ K. We use their derived properties,
after rescaling them to the IMF and SFR(IR) calibrations adopted in this paper (see Sect.\ \ref{s_obs}).
To this we add 11 measurements (6 detections, 5 upper limits) of galaxies between $z=6.0$ and 9.11 from
\citeharikane, who report three new observations and eight others not included in the \cite{Matthee2019Resolved-UV-and} 
compilation. For consistency, we used the SFR(UV) and SFR(IR) values, and we carefully rescaled their results
to a single, consistent IMF and to the same SFR(IR) calibration.  
Finally, for sources that are not detected in the dust continuum, we corrected for hidden SF by applying the
IRX--$\beta$ relation derived at $z \sim 5.5$ from the ALPINE sample when the UV slope is reported. 
For the majority of the galaxies, the correction turns out to be small, since their UV slope is fairly blue.
The data are plotted Fig.\ 5.

Interestingly, most of the known $z \ga 6$ galaxies largely follow the same behaviour/trend as the ALPINE galaxies.
At SFR(tot)$\ga 30 $\msunyr,\ very few points deviate by more than $1 \sigma$ from the \citedelooze\ relation.
Differences between our results and those shown in  \citeharikane\ are explained by several effects: 
by our use of a single IMF consistent with the \citedelooze\ relation, a consistent use of calibrated 
SFR determinations for all sources (no SED-based SFR as for some of their sources) following 
\cite{Matthee2019Resolved-UV-and}, and finally by the adoption of conservative line widths to determine
upper limits on \lcii. 

\cii\ is undetected in a significant number of galaxies at SFR$\la 30$\msunyr.
Assuming a FHWM$=150$ \kms\ for the ALPINE sources and the 15 upper limits from the 
$z>6$ data discussed above, we find that the deepest $3 \sigma$ upper limits are $\log(\lcii/\lsun) <7.8$
for the bulk of the data. While a fraction of those are well within the scatter around the `local' \citedelooze\ 
relation, several are probably below this, which pushes the average \lcii/SFR ratio of both the ALPINE sample
and the full high-$z$ galaxy sample to $\log(\lcii/)$SFR$\approx 6.85$ \lsun/\msunyr,  approximately
0.2 (0.1) dex lower than the reference value from \citedelooze\ for the \hii/starburst (complete) galaxy samples
(see also Fig.\ \ref{fig_cii_lya}).
Two non-detections at SFR$< 30$\msunyr\  are strongly underluminous in \cii\ compared to
the rest of the sample: these 
are two lensed galaxies at $z>8$, A2744-YD4 at $z=8.382$ and MACS1149-JD1 at $z=9.11$ observed 
by \cite{Laporte2019The-absence-of-}, which were previously detected by ALMA in the 
[O~{\sc iii}] 88$\mu$m line by \cite{Laporte2017Dust-in-the-Rei} and \cite{Hashimoto2018The-onset-of-st}.
On the other hand, another [O~{\sc iii}]-detected lensed $z=8.312$ galaxy 
\citep[MACS0416-Y1 from][]{Tamura2019Detection-of-th} is detected in \cii\ and follows the observed
trend well.

To quantify this behaviour again, we use the Bayesian fit including censored data.
The results including the uncertainties and upper limits on \lcii\ are shown by the green lines/band in Fig.\ 6.
The fit shows that the inclusion of the upper limits primarily leads to a somewhat steeper (super-linear) slope
($1.28 \pm 0.1$) in the \lcii--SFR relation for galaxies at $z > 4,$ and to an overall (but slight) decrease of the normalisation,
for instance\ to a lower \cii\ luminosity on average at a given SFR, as already mentioned and shown in Fig.\ \ref{fig_cii_lya}.
Overall, the observational data shows a behaviour that is quite comparable to the mean \lcii--SFR relations
predicted by the models of \cite{Lagache2018The-CII-158-mum} between $z=4$ and $z=6$.
The fit to the data shown in Fig.\ \ref{fig_cii_z69} is also similar to the mean relation obtained from the 
recent simulations of high-$z$ galaxies by \cite{Arata2020Starbursting-O-}, who find a somewhat steeper slope of 1.47.
In contrast, the cosmological plus radiative transfer simulations of \cite{Leung2020Predictions-of-} for $z \sim 6$ 
galaxies predict a \lcii--SFR relation with, on average, significantly lower \cii\ luminosities than the observational data 
shown here. Furthermore, the slope of their best-fit relation ($0.66 \pm 0.01$) is flatter than unity and than
those found in our study.

Interestingly, with the enlarged sample (ALPINE and $z>6$ galaxies) our fits yield slopes steeper than unity
using both options for the \cii\ upper limits (agressive versus conservative; cf.\ above). This result is mostly
driven by a few additional data points at low \lcii\ and low SFR, which have low uncertainties on \lcii\ and thus a
fairly strong leverage. Whether these points are truly representative of the bulk of the population of
fainter galaxies or `outliers' remains to be confirmed with new observations probing this regime.

Overall, we conclude that the \cii\ measurements (detections and upper limits) of star-forming galaxies 
at $z \sim 4 - 8$ follow a unique relation between \lcii\ and SFR(UV)+SFR(IR) quite well over nearly two orders
of magnitude in the \cii\ luminosity. This holds for a wide variety of galaxy types (LAEs, LBGs primarily, plus two SMGs from
\cite{Marrone2018Galaxy-growth-i} included in the \citeharikane\ compilation) issued from different selections.
Taking the \cii\ non-detections into account, the \lcii--SFR rrelation  appears to be somewhat steeper and offset from 
the `local' counterpart determined by  \citedelooze. 
Whether the relation shows a turnover below SFR$\la 10-30$ \msunyr, as suggested, for example\ by \cite{Matthee2019Resolved-UV-and}, cannot be established from the available data. This would
require more sensitive measurements.

\subsection{Is there a dependence of \cii\ with the \lya\ equivalent width in high-$z$ galaxies?}
Several authors have pointed out that galaxies with an increasing \lya\ equivalent width 
show a fainter \cii\ emission, compared to expectations from the local \cii\--SFR relation
\citep[see e.g.][]{Carniani2018Kiloparsec-scal,Harikane2018SILVERRUSH.-V.-,Harikane2019Large-Populatio}.
They relate this trend to a possible increase of the ionisation parameter or to a low covering fraction 
of photodissociation regions with increasing \lya\ emission.
Benefitting now from the large amount of new data from ALPINE for which we also have EW(\lya) measurements
(for 58 \cii-detected plus 33 non-detected sources, taken from \citecassata), we show the behaviour of \lcii/SFR(tot) 
as a function of EW(\lya) in Fig.\ \ref{fig_cii_lya}. The ALPINE sources cover a wide range of \lya\ equivalent widths, also including relatively 
large EWs, some of which were selected as LAE. Again, a large scatter is found in \lcii/SFR at all values of EW(\lya),
and the fit to all the data including the non-detections shows a weak dependence of EW(\lya) on \lcii/SFR\footnote{
The Bayesian fit yields $\log(\lcii/{\rm SFR}) = (6.99 \pm 0.07) -(0.10 \pm 0.06) \times \log(EW)$ in the units plotted in Fig.\ \ref{fig_cii_lya}.}.
Clearly, we cannot claim a strong anti-correlation of \cii\ with increasing \lya\ equivalence width, in contrast to  \citeharikane.
The difference compared to their work lies in our significantly larger dataset,  our use of `uniformised' values for SFR, 
and the use of a more conservative line width for the determination of the upper limits on \cii\ luminosities, as already mentioned
above.

\begin{figure*}[tb]
{\centering
\includegraphics[width=9cm]{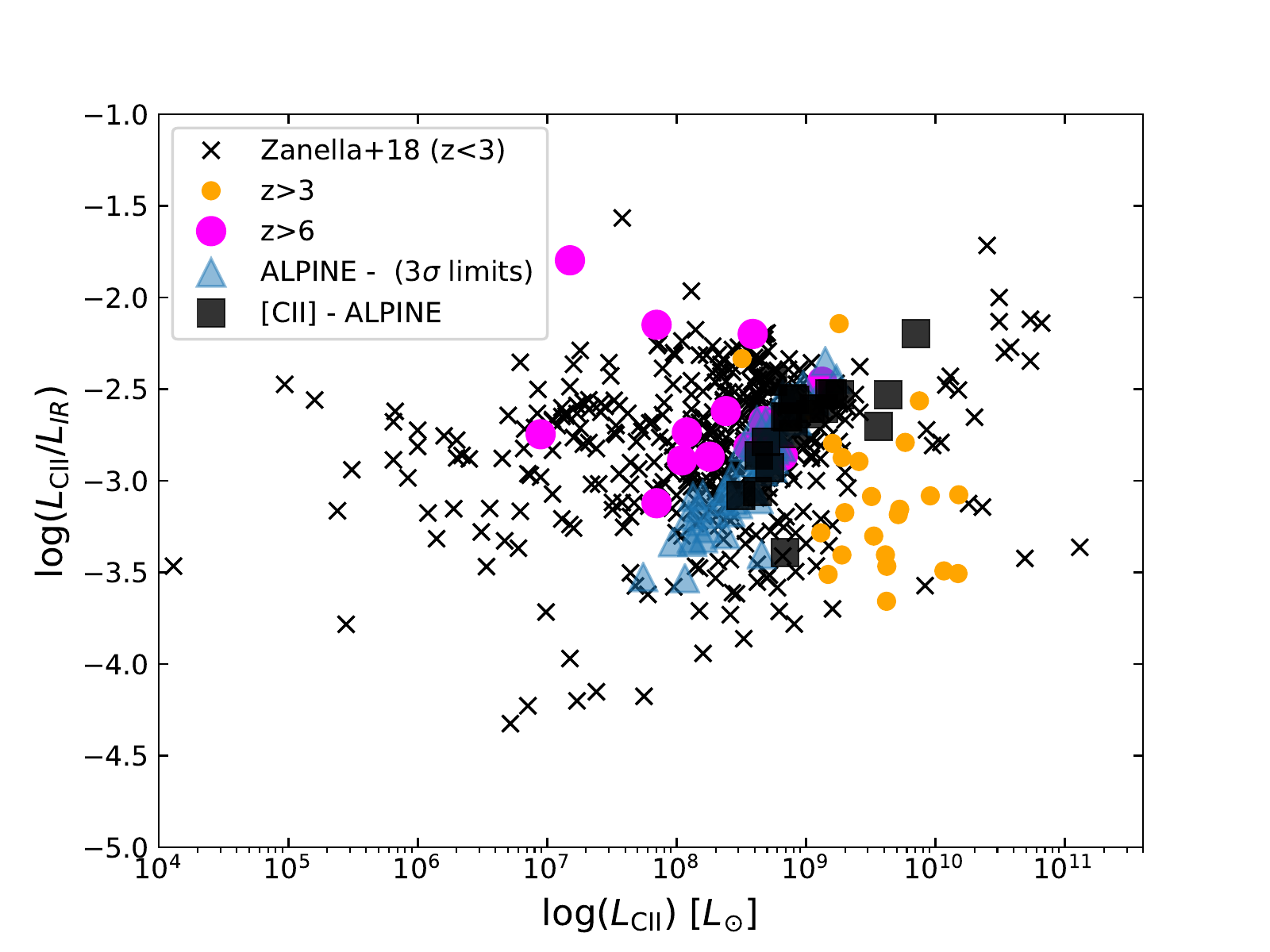}
\includegraphics[width=9cm]{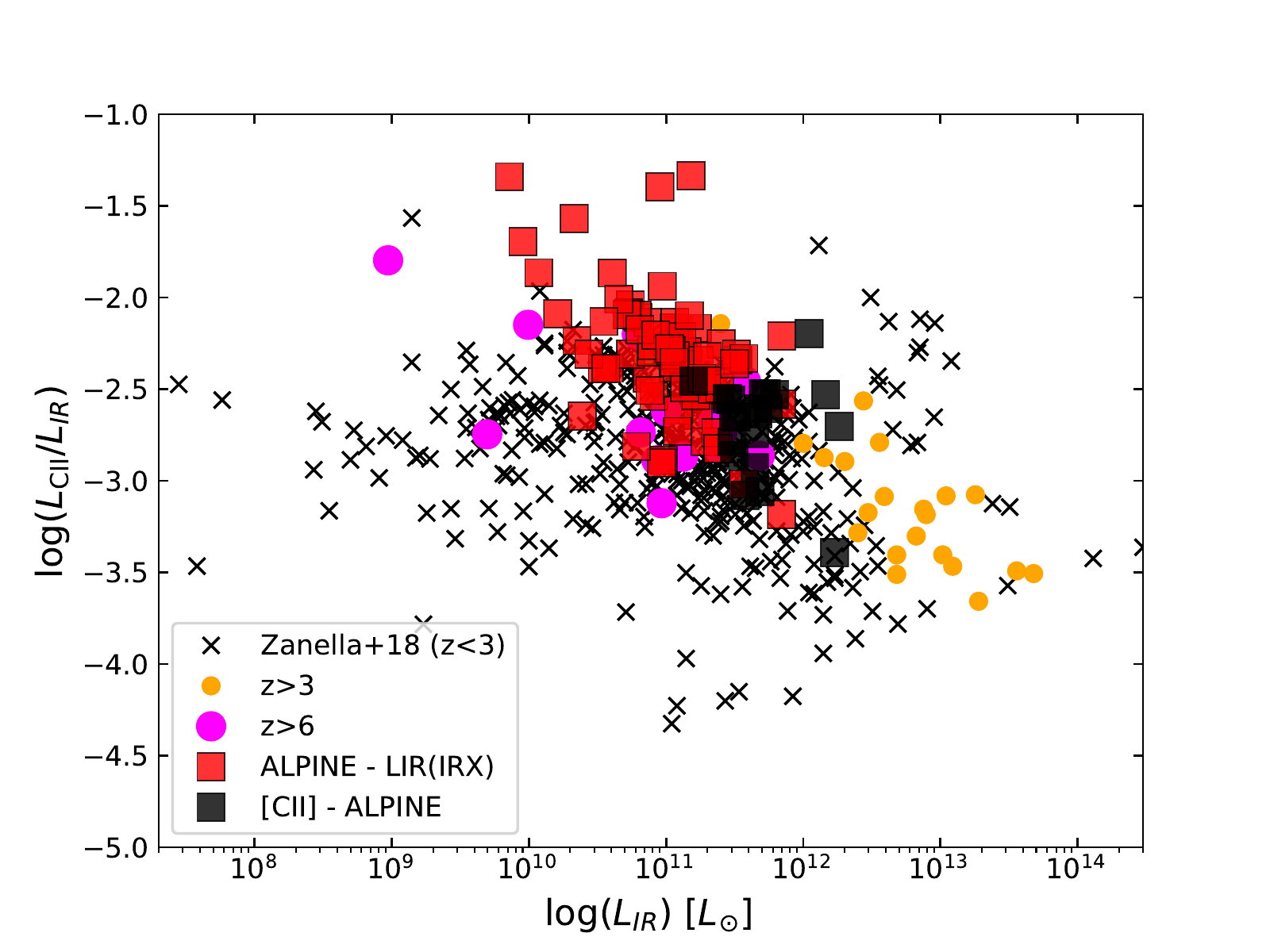}
\caption{\lcii/\lir\ versus \lcii\ (left panel) and \lir\ (right) for the ALPINE sources and comparison samples where the \cii\ line is detected. 
For the $z>6$ sources, taken from the \cite{Matthee2019Resolved-UV-and} sample, we adopt $T_d=45$ K 
for the detections and \lir\ upper limits. 
For galaxies not detected in the dust-continuum, we show their  \lcii/\lir\ 3 $\sigma$ lower limit in the left panel as triangles.
In the right panel, the \lir\ limits of the ALPINE sources have been replaced by the \lir\ values computed from the 
IRX--$\beta$ relation, where possible.}
}
\label{fig_cii_ir}
\end{figure*}

\section{Observed \Cii\ line to IR continuum ratios}
\label{s_cii_ir}
With the exception of some lensed sources from the SPT survey \citep{Gullberg2015The-nature-of-t},
the $z>3$ galaxies currently detected in the dust continuum have typical (lensing-corrected, if applicable) IR luminosities 
in the range of $\lir \ga 10^{11}$  to $2 \times 10^{12}$ \lsun, hence they are LIRG or ULIRG by definition.
This is also the case for the continuum-detected ALPINE sources (see \citedata).
In this regime of high-IR luminosities, low-$z$ galaxies show the well-known `\cii\  deficit', meaning\ a drop
of \lcii/\lir\ towards high \lir\ \citep[see e.g.][]{Malhotra2001Far-Infrared-Sp,Gracia-Carpio2011Far-infrared-Li}.
It is therefore of interest to examine how high redshift galaxies, the ALPINE sample in particular, and others,
behave in this respect.

Since the IR continuum is undetected in many observations of normal star-forming galaxies at
high redshift, we first plot the \lcii/\lir\ ratio as a function of the \cii\ luminosity instead of \lir.
The result is shown in the left panel of Fig.\ 8.
where we show the data for all the \cii-detected galaxies
of ALPINE, the $z >6$ data from the compilation of \cite{Matthee2019Resolved-UV-and}, 
other \cii\  detections (non-AGN-dominated sources) at $z>3$ taken from the compilation in \cite{Gullberg2015The-nature-of-t},
the SPT sources of \cite{Gullberg2015The-nature-of-t}, and observations at $z<3$ from the compilation of 
\cite{Zanella2018The-C-ii-emissi}. We note that we use the total IR luminosity here (from 8-1000 \micron), following \cite{Zanella2018The-C-ii-emissi}, for example, whereas other authors use the far-IR luminosity, $L_{\rm FIR}$ (from 40-122 \micron\ restframe), as a reference; 
in the \cite{Zanella2018The-C-ii-emissi} compilation, one typically has \lir/$L_{\rm FIR}=1.6$. 
For the SED template used for ALPINE one has \lir/$L_{\rm FIR}=1.628$.

The ALPINE sources detected in the continuum show a ratio \lcii /\lir $\sim (1-3) \times 10^{-3}$  (or a factor of 1.628 higher when compared
to $L_{\rm FIR}$), comparable to the `normal' $z<1$ sources, whereas the IR luminous \cii-deficient galaxies have
 \lcii/\lir $< 10^{-3}$. The same is also found for the other continuum-detected $z>6$ galaxies, and the majority of the
$z \sim 4-7$ sources that are currently undetected in the dust continuum are also compatible with normal or higher \lcii /\lir\ ratios.
In other words, the majority of the $z >4$ galaxies where \cii\ is detected do not seem to show a deficit in \lcii/\lir, similar to earlier findings at
lower redshift  \citep[e.g. $z \sim 1-2$, ][]{Zanella2018The-C-ii-emissi}.
On the other hand, the SPT sample, which is significantly brighter than the ALPINE sources and the 
$z >6$ LBGs and LAEs, shows several sources with \lcii/\lir $< 10^{-3}$ and an increasing \cii-deficit 
at IR luminosities above $\ga 10^{12}$ \lsun, as shown by \cite{Gullberg2015The-nature-of-t}. One may therefore
speculate that a \cii\  deficit is also present in high-$z$ galaxies, albeit at intrinsically higher IR luminosities,
again suggesting that the \cii/IR-deficit is not a universal property, as already suggested earlier 
\citep[cf.][]{Zanella2018The-C-ii-emissi}.
 
In the right panel of Fig.\ 8
we show a more classical version of the dependence of the \lcii /\lir ratio,
plotted as a function of \lir, where the IR luminosity of the high-$z$ ($z \sim 4-6$ galaxies from ALPINE and the $z>6$ sample)
is taken from the observations or has been computed from the ALPINE IRX--$\beta$ relation (as used in Figs.\ 4--7) 
for the continuum non-detected sources. Most of the latter sources have predicted $\lir \sim 10^{10} - (3 \times 10^{11})$ \lsun,
and \lcii /\lir\ ratio ranges between $10^{-3}$ and $10^{-2}$, compatible with the bulk of the $z<3$ galaxies.

Since the observed decrease of  \lcii/\lir\ in low and high-$z$ galaxies is known to correlate with the increasing dust temperature
\citep[e.g.][]{Malhotra2001Far-Infrared-Sp,Gracia-Carpio2011Far-infrared-Li,Diaz-Santos2013Explaining-the-,Gullberg2015The-nature-of-t}, 
one might be tempted to conclude that the `normal'  \lcii/\lir\ ratio found for the majority of the ALPINE galaxies and $z>6$ LBGs and LAEs,
could indicate that these sources do not harbor particularly hot dust.
In the context of the intensely debated uncertainties on the typical dust temperatures of normal galaxies in the early Universe
\citep[see e.g.][]{Bouwens2016ALMA-Spectrosco,Faisst2017Are-High-redshi,Ferrara2017The-infrared-da}, this would have
important implications. 
Conversely, from the analysis of \cite{Magdis2014A-Far-Infrared-} one would expect some \cii\ deficiency for the ALPINE continuum-detected
galaxies at $T_d \sim 42$ K of our template, but this effect is not seen in our data.
In any case, Fig.\ 8
should not be over-interpreted since the inferred IR luminosity itself depends
on the assumed IR SED template, meaning\ directly or indirectly on $T_d$. Independent constraints on the IR SED and dust temperature
of high-$z$ galaxies are clearly needed.

\section{Discussion}
\label{s_discuss}

If star formation were unobscured in most of the $z \sim 4-6$ galaxies covered by the ALPINE survey, our observations would indicate that \cii\
is over-luminous at a given SFR$\approx$SFR(UV), compared to the observed correlation for low redshift galaxies (see Fig.\ \ref{fig_cii_sfr}).
At face value, such a conclusion would be quite in contrast with earlier studies of $z>6$ galaxies, which for example
have argued that \cii\ was less luminous than expected from comparisons with the low-$z$ reference sample
\citep[see e.g.][]{Ouchi2013An-Intensely-St,Bradac2017ALMA-C-II-158-m,Harikane2018SILVERRUSH.-V.-}.
 
However, as argued above, it seems much more likely that a fraction of the UV light from star formation is attenuated by dust,
in the majority of our targets, as well as in those from which we do not detect dust-continuum emission with ALMA.
Indeed, a relatively small correction of  SFR(UV) -- upward by a factor of $\sim 2$ on average -- is sufficient to bring the \cii\
measurements on average into agreement with the local \lcii--SFR relation (cf.\ Sect.\ \ref{s_hidden}). 
The amount of this correction appears very reasonable from several points of view: 
firstly, it corresponds to the average correction obtained from multi-band SED fitting of the rest-UV-to-optical
SED;  and secondly, the same correction is found, on average, by applying an empirically calibrated IRX--$\beta$ relation
of the ALPINE galaxies derived from stacking to the individual ALPINE galaxies, which are not detected in the dust continuum.
Finally, stacking the continuum in bins of \lcii\ also indicates a necessary correction to the SFR(UV) as shown by \citedata, 
leading to a fair agreement of the stacked data with the local relation.

Taking into account a correction for dust-obscured star formation, we then examined and derived the 
empirical relation between \lcii\ and the total SFR(tot) for $z>4$ galaxies, using both the ALPINE sample 
covering $z \sim 4-6$ and data from the literature for $z \sim 6-9$ galaxies. We also included \cii\ non-detections
in a Bayesian linear fit of the data (Sect.\ \ref{s_universal} and Appendix).
We also stress the importance of a consistent use of SFR calibrations, IMF normalisations, 
and empirically motivated \cii\ line widths to compute upper limits \citep[see also][]{Matthee2019Resolved-UV-and}, 
which must be taken into account for meaningful and consistent comparisons of different datasets and to establish, 
for example, a possible evolution of the \lcii--SFR relation with redshift.
Some of our results are obviously also subject to uncertainties and future improvements, which we now briefly discuss.

Making reasonable assumptions on the dust-obscured SFR and using for the first time a large sample of up to 150 galaxies,  
we have shown that the \cii\ luminosity of high-$z$ ($z>4$)  galaxies correlates well with the total SFR,  
over approximately two orders of magnitude in SFR.
The data is described well by a linear relationship between $\log(\lcii)$ and $\log({\rm SFR_{tot}})$ with a slope
close to unity ($b \sim 0.8-1.3$) (see Table \ref{tab_fits}). However, the exact slope of the relation depends in part on the \cii-undetected sources,
and hence on the detailed assumptions on the upper limits, which depends not only on assumed line widths, 
but also on the hypothesis about size (point-like or slightly  extended sources).
Deeper observations for some of the ALPINE targets would be easily attainable with ALMA, and helpful to better understand
the sources with $\log(\lcii) \la 10^8$ \lsun. 
To firm up the result of a possibly steeper \lcii--SFR relation at high-$z$ than for local galaxies, it
is also clearly important to acquire more measurements of fainter galaxies with lower star formation rates,
ideally at SFR $\la 1-3$ \msunyr, where currently only very few observations of lensed galaxies have been
obtained \citep{Knudsen2016C-ii-emission-i,Bradac2017ALMA-C-II-158-m}.

Although we fitted the available data with simple linear relations (i.e.\ a power-law dependence of \lcii\ on SFR),
nature may be more complicated, and the conditions may be different in high-redshift galaxies.
The high-$z$ data discussed here do not allow us to exclude different behaviour at low SFR or low \lcii,
as suggested, for example,\ by  \cite{Matthee2019Resolved-UV-and}.
However, on resolved scales in our Galaxy and for individual galaxies from the nearby Universe up to $z \sim 1-3$,
different studies have empirically established a correlation between \cii\ and the total SFR
with simple power laws with exponents of $\sim 0.8 - 1.2$ extending over approximately six orders of magnitude
\citep[see e.g.][]{Pineda2014A-Herschel-C-II,2011MNRAS.416.2712D,De-Looze2014The-applicabili,Zanella2018The-C-ii-emissi},
and which include the range probed by high-$z$ observations. 
From an empirical point of view and in the absence of strongly deviating data, we did not consider
other functional forms of the \cii--SFR relation.

Beyond \cii, the second fundamental quantity for this work is obviously the total SFR, which is currently
not easy to determine, due to technical limitations (insufficient sensitivity to detect dust-continuum emission)
and our limited knowledge of the dust properties and IR template, which are required to infer the
total IR luminosity, and hence the dust-obscured part SFR(IR). On the other hand, SFR(UV) is easy
to determine for the galaxies of interest here, since all of them were previously detected at these
wavelengths for our survey (\citesurvey). The IR template used in our work to translate the rest-frame
158 \micron\ continuum measurements into the total \lir\ has a similar `bolometric correction' to a 
modified black body (MBB) with $T_d \approx 42$ K (\citedata). Using, for example, the empirical template of 
\cite{Schreiber2018Dust-temperatur} would imply \lir\ values that are higher by 43\%, comparable to an MBB with
$T_d \sim 45$ K. If even higher dust temperatures were appropriate, \lir\ would, for example, increase by a factor of
1.87 (3.79) for $T_d =50$ (60) K compared to  (\citedata). 

With our assumptions and the adopted IRX--$\beta$ correction, for the ALPINE galaxies
one has SFR(UV) $\approx$ SFR(IR) on average, and most galaxies have SFR(UV) $\la 2$ SFR(IR).
In this case, an increase of \lir\ by a factor of two (3) would translate as an increase of the total SFR
by a factor of 1.5--1.6 (2--2.3).
This effect could thus shift the \cii--SFR relation by this amount, away from the local relation.
Whether (and by how much) this could also change the slope of the relation depends if the dust-obscured SFR 
fraction is constant in all galaxies, and how the dust temperature may vary with galaxy properties,
all of which are largely unknown for high-$z$ galaxies.

Clearly, accurately determining the total SFR of high-$z$ galaxies will lead to significantly
more robust results on the \cii--SFR relation in the distant Universe. Efforts are under way to
constrain the dust temperatures at high-$z$ 
\citep[e.g.][]{Hirashita2017Dust-masses-of-,Faisst2017Are-High-redshi,Bakx2020ALMA-uncovers-t}.
Alternatively, the JWST should soon provide measurements of rest-optical lines including hydrogen 
recombination lines, which will allow one to determine, for example, the \ha\ SFR and dust corrections
using the Balmer decrement for high-$z$ galaxies. This could become an important and complementary
method to nail down some of the uncertainties discussed here, and indirectly also to constrain 
the dust temperature and \lir\ of distant galaxies.

Finally, we would like to caution that the \cii\ luminosity may not necessarily trace
the SFR accurately in general, especially in high redshift galaxies.
Although \lcii\ empirically\ correlates well with the SFR, the main physical reason(s) for 
this dependence are not well understood and predictive models are therefore difficult to construct,
presumably largely since \cii\ is known to originate from a broad range of ISM phases and regions 
with different conditions \citep[see e.g.][]{Vallini2015On-the-CII-SFR-,Lagache2018The-CII-158-mum,Ferrara2019A-physical-mode,Popping2019The-art-of-mode}.
In fact, the empirical correlations of \lcii\ determined here and in earlier studies are with the
UV+IR luminosity, or a combination of the two, which can be converted to the SFR if one assumes 
a particular star formation history and age of the population. More fundamentally, the data thus
probably indicate a correlation of the \cii\ luminosity with the intrinsic UV luminosity of the galaxy
-- part of which emerges in the UV, and the other part after processing by dust in the IR --
which is also physically significant, since \cii\ requires photons capable of singly
ionising carbon atoms, that is to say\ with energies $>11.26$ eV (wavelength $< 1102$ \AA).
This implies in particular that \lcii\ does not need to closely follow the instantaneous SFR
in galaxies with strongly varying (irregular, burst, etc.) star formation histories, where 
significant variations between \luv\ and the SFR are expected \citep[see e.g.][]{schaerer2013,Madau2014Cosmic-Star-For}.
Such situations  are probably more  common in the early Universe, and one may therefore
expect a better correlation of \lcii\ with the intrinsic (total) UV luminosity than with other
tracers of the SFR, such as H recombination lines.

Furthermore, \cii\ emission may also depend on metallicity and other galaxy properties
such as the gas fraction, distance from the main sequence, and so on
\citep[cf.][]{Vallini2015On-the-CII-SFR-,Lagache2018The-CII-158-mum,Zanella2018The-C-ii-emissi}.
Currently, these quantities are largely unknown for high-$z$ galaxies and are difficult to measure.
In addition, the ALPINE sample is, by construction, not well placed to examine possible dependences on the main sequence
distance, since it selected main sequence galaxies and subtle variations from it are difficult to measure.
Future independent measurements may yield a more refined picture of the main processes governing the emission
of \cii\ in distant galaxies.

Other results from the rich ALPINE dataset are presented elsewhere. Beyond those already mentioned
earlier, these cover, for example, the use of \cii\ to estimate the amount of gas in galaxies, morphological studies, 
the detection of \cii\ in the circumgalactic medium, studies of the kinematics between \cii\ and \lya, the discovery
of very obscured sources, and several other topics.
For more details on these issues, see \cite{Dessauges-Zavadsky2020The-ALPINE-ALMA,Fujimoto2020The-ALPINE-ALMA,
Ginolfi2020The-ALPINE-ALMA,Cassata2020The-ALPINE-ALMA,Romano2020The-ALPINE-ALMA}.

\section{Conclusions}
\label{s_conclude}

We analysed the new \Cii\ measurements from the ALPINE  survey of star-forming galaxies at $z \sim 4-6$ 
(\citesurvey, \citedata, \citeancillary), which for the 
first time provides a large sample (118 galaxies) to study \cii\ emission and its correlation with the star formation rate at high redshift. 
We examined whether our data and other observations at $z >6$ -- now totalling 153 galaxies -- 
are compatible with the observed correlation between the \cii\ luminosity and SFR found at lower redshift, and described in the \citedelooze\ 
reference sample.

To compare the high-$z$ observations to the earlier data, we used consistent SFR calibrations (based on UV and IR continuum luminosities) 
and a carefully homogenised IMF. We also took into account the \cii\ non-detections, which are translated into upper limits on \lcii\ adopting 
empirically motivated assumptions on the \cii\ line widths, which we re-examined using our own data and literature data  (see Fig.\ 1).
 %
The ALPINE galaxies, which are both detected in \cii\ and the dust continuum, show a good agreement with  
the low-$z$ \lcii--SFR relation when considering the total SFR(UV+IR). A fraction of the non-detected ALPINE galaxies in the dust continuum 
appear over-luminous in  \lcii\ compared to expectations from the \citedelooze\ relation, when no
correction for dust attenuation is made (see Fig.\ \ref{fig_cii_sfr}). 
This is in contrast with earlier studies, which have often reported apparent deficits of \cii\ in high-$z$
galaxies \citep[e.g.][]{Ouchi2013An-Intensely-St,Inoue2016Detection-of-an,Harikane2018SILVERRUSH.-V.-}.
Using the results from two different stacking methods, described in \citedata, \cite{Fudamoto2020The-ALPINE-ALMA}, and SED fits allows us to 
account for dust-obscured star formation in these galaxies, thus increasing their total SFR by a factor of $\sim 2$ on average,
which brings the ALPINE galaxies into agreement  with the local \cii--SFR relation (Fig. 3, right, and Fig. 4).

When conservative upper limits from the \cii\ non-detected galaxies ($\sim 1/3$ of the ALPINE survey) are also considered, 
we find that \lcii\ scales linearly with the total SFR for the ALPINE sample, although with a slightly lower normalisation
(\lcii) than the local \hii/starburst galaxy sample of \citedelooze.
Using more agressive upper limits leads to a steepening of the \lcii--SFR relation.
A steeper increase of \lcii\ with SFR is also found when all the available \cii\ measurements (detections and upper limits) 
at $z \sim 4-8$, including other ALMA measurements from the literature are combined (Fig.\ 6).
Given the remaining uncertainties on the \cii\ non-detected galaxies and the exact amount of dust-obscured SFR,
we conclude that the exact slope of the \lcii--SFR relation at $z>4$ is not firmly established.

Upon analysing the homogenised sample of 153 $z>4$ galaxies with \cii\ measurements (detections or upper limits), we find that very few galaxies 
deviate significantly from the bulk of the sample,
and that most $z \sim 4-8$ galaxies show an \lcii--SFR relation that is not very different from that of low-$z$ galaxies nearly 13 Gyr later.
In other words, the currently available data show no strong evidence for a deficit of \cii\ from $z \sim 4$ to 8,
in contrast to several earlier results, but in line with other suggestions \citep{Carniani2018Kiloparsec-scal,Matthee2019Resolved-UV-and}.
The only strong outliers from the \lcii--SFR relation are two galaxies at $z >8$  with [O~{\sc iii}] 88$\mu$m line detections
with ALMA and no \Cii\ \citep{Laporte2019The-absence-of-}, which may indicate a more fundamental change of properties
in the very early Universe.
 
We also examined the behaviour of \lcii/SFR with the observed \lya\ equivalent width of the ALPINE galaxies and
literature data, and we do not find a strong dependence of the \cii\ excess or deficiency with EW(\lya) at $z>4$ (Fig.\ \ref{fig_cii_lya}), 
in contrast with earlier suggestions \citep[e.g.][]{Harikane2018SILVERRUSH.-V.-,Harikane2019Large-Populatio,Matthee2019Resolved-UV-and}.
Finally, we show that the derived ratio \lcii /\lir $\sim (1-3) \times 10^{-3}$  for the ALPINE sources, comparable to that
of `normal' galaxies at lower redshift (Fig.\ 8).

Overall, our results, using 153 galaxies at $z>4$, suggest that the \cii\ luminosity can be used to trace the SFR at these
high redshifts, although the scatter is higher than at low redshift, as already indicated\ by \cite{Carniani2018Kiloparsec-scal}, for example.
Furthermore, there is some evidence for a possible steepening of the \lcii--SFR relation compared to $z < 3$, although this
needs to be confirmed with future measurements and better constraints on dust-obscured star formation in high-$z$ galaxies,
which can be obtained with new ALMA and future JWST observations.

\begin{appendix}
\section{Fits for the \cii--SFR relation at high redshift}
The ALPINE dataset and the data for $z>6$ galaxies from the literature were fitted  
using a Bayesian fit including censored data following the method of \cite{Kelly2007Some-Aspects-of},
which is implemented in the  {\em linmix} python package.
In Table \ref{tab_fits}, we list the resulting fit coefficients of the linear fits of the form
$\log(\lcii/\lsun) = a + b \times \log({\rm SFR_{tot}}/\msunyr)$ and their uncertainties
obtained for different combinations of datasets, assumptions on SFR(tot), and adopted \cii\ upper limits.
Not all combinations are shown and discussed in the text; those shown in Figures are indicated in the 
last column in the table.

 \begin{table*}[htb]
\caption{Fit coefficients from Bayesian fits 
including censored data: $(a,b)$=(offset, slope) and their uncertainties (standard deviation).
Col.\ 1 indicates the dataset used, col.\ 2 the total SFR used, col.\ 3 the \cii\ limits.
Col. 8 indicates the figure number showing the corresponding data and fit in some cases.}
\begin{center}
\begin{tabular}{llllllll}
Dataset & SFR & \cii\ limits & offset & std(offset) & slope & std(slope) & Fig. \\
\hline \\
ALPINE & UV+IR &  3-$\sigma$ limits            & 7.03 &  0.17 & 1.00 & 0.12 & 2 \\  
ALPINE & UV+IR &  6-$\sigma$        & 7.37 &  0.14 & 0.83 & 0.10 \\ 
ALPINE & SED & 3-$\sigma$                & 7.09 &  0.21 & 0.84 & 0.13 & 3 right \\ 
ALPINE & SED & 6-$\sigma$            & 7.43 &  0.17 & 0.70 & 0.10 &  \\ 
ALPINE & UV+IRX & 3-$\sigma$              & 6.61 &  0.20 & 1.17 & 0.12 & 4 \\ 
ALPINE &UV+IRX & 6-$\sigma$               & 7.05 &  0.15 & 0.96 & 0.09 &  5 \\ 
ALPINE+$z>6$ & UV+IRX & 3-$\sigma$     & 6.43 &  0.16 & 1.28 & 0.10 & 6 \\ 
ALPINE+$z>6$ & UV+ (IRX for ALPINE only)& 3-$\sigma$     & 6.51 &  0.15 & 1.24 & 0.10 &  \\ 
ALPINE+$z>6$ & UV+IRX & 6-$\sigma$ & 6.66 &  0.14 & 1.17 & 0.09 \\ 
\end{tabular}
\end{center}
\label{tab_fits}
\end{table*}%

\end{appendix}

\begin{acknowledgements}
This paper is based on data obtained with the ALMA Observatory, under Large Program 2017.1.00428.L. ALMA is a partnership of ESO (representing its member states), NSF(USA) and NINS (Japan), together with NRC (Canada), MOST and ASIAA (Taiwan), and KASI (Republic of Korea), in cooperation with the Republic of Chile. The Joint ALMA Observatory is operated by ESO, AUI/NRAO and NAOJ. 
DS, MG and MD  acknowledge support from the Swiss National Science Foundation.
AC, CG, FL, FP and MT acknowledge the support from grant PRIN MIUR 2017 - 20173ML3WW\_001.
EI acknowledges partial support from FONDECYT through grant N$^\circ$\,1171710.
GCJ and RM acknowledge ERC Advanced Grant 695671 ``QUENCH'' and support by the Science and Technology Facilities Council (STFC).
GL acknowledges support from the European Research Council (ERC) under the European Union's Horizon 2020 research and innovation programme (project CONCERTO, grant agreement No 788212) and from the Excellence Initiative of Aix-Marseille University-A*Midex, a French ``Investissements d'Avenir'' programme.
DR acknowledges support from the National Science Foundation under grant numbers AST-1614213 and AST-1910107 and from the Alexander von Humboldt Foundation through a Humboldt Research Fellowship for Experienced Researchers.
ST acknowledges support from the ERC Consolidator Grant funding scheme (project ConTExT, grant No. 648179). The Cosmic DAWN Center is funded by the Danish National Research Foundation under grant No. 140
LV acknowledges funding from the European Union's Horizon 2020 research and innovation program under the Marie Sklodowska-Curie Grant agreement No.\ 746119.

\end{acknowledgements}
\bibliographystyle{aa}
\bibliography{merge_misc_highz_literature}

\end{document}